\documentclass[sglanonrev]{informs4_workingpaper}

\usepackage{amsmath}
\usepackage{amssymb}
\usepackage{diagbox}
\usepackage{multirow}
\usepackage{bm}
\usepackage{booktabs}
\usepackage{subcaption}
\RequirePackage{endnotes}

\usepackage{xurl}
\usepackage[hidelinks]{hyperref}

\usepackage{enumitem}
\usepackage[linesnumbered,ruled]{algorithm2e}
\usepackage{soul}

\usepackage{placeins}

\usepackage{xcolor}

\usepackage{natbib}
\bibpunct[, ]{(}{)}{,}{a}{}{,}%
\def\bibfont{\small}%

\EquationsNumberedThrough
\TheoremsNumberedThrough
\ECRepeatTheorems

\MANUSCRIPTNO{}

\begin{document}

\RUNAUTHOR{Pauphilet and Wu}
\RUNTITLE{Parallel Node Generation and Exploration Algorithm on GPU}
\TITLE{PaNGEA: Parallel Node Generation and Exploration Algorithm on GPU}

\ARTICLEAUTHORS{%
\AUTHOR{Jean Pauphilet}
\AFF{London Business School, \EMAIL{jpauphilet@london.edu}}

\AUTHOR{Yupeng Wu}
\AFF{London Business School, \EMAIL{yupengw@london.edu}}
}

\ABSTRACT{%
Primal heuristics for finding high-quality feasible solutions are an important component in mixed-integer optimization (MIO) solvers. Recent advances in GPU-accelerated optimization algorithms show the potential of GPU acceleration for continuous optimization. In this paper, we introduce the Parallel Node Generation and Exploration Algorithm (PaNGEA), a GPU-friendly MIO primal heuristic. PaNGEA explores restricted subproblems by combining linear-relaxation solves with a local-search procedure designed for efficient batched execution on GPUs. 
In addition, instead of relying on a single heuristic for iterative variable fixing, we leverage GPU batching capabilities to generate and explore multiple subproblems in parallel. 
PaNGEA leverages GPU capabilities in two ways. 
First, on 283 MIPcc26 and MIPLIB instances, implementing a single-node primal heuristic on the GPU reduces the average gap integral by 8--18\% relative to its CPU counterpart. Second, generating and exploring multiple nodes in parallel further reduces the average gap integral by 12--19\%.}


\maketitle

\section{Introduction}
Mixed-Integer Optimization (MIO) is a powerful paradigm for modeling and solving real-world decision problems. 
While NP-hard in the worst case \citep{Wolsey2008MIO}, many large-scale MIO problems can now be solved efficiently, with speed-ups attributable to both better algorithms and hardware \citep{bixby2012brief}.

On the algorithm side, efficient solvers combine the implementation of certifiably optimal methods like branch-and-bound with efficient primal heuristics. By finding high-quality solutions early, primal heuristics can help prune the branch-and-bound tree quickly. Conversely, they benefit from being applied on restricted subspaces such as intermediate nodes of the branch-and-bound process \citep{fischetti2003local,danna2005exploring}.

On the hardware side, current MIO solvers have mostly benefited from the improvement in CPU chips. Over the past 10 years, GPUs have received a lot of attention from the deep learning community \citep{krizhevsky2012imagenet}. Their ability to solve multiple linear systems in parallel creates new opportunities for optimization algorithms. For example, \citet{applegate2023faster,lu2025cupdlp} propose efficient GPU implementations of first-order methods for linear (continuous) optimization problems.
However, exploiting GPU parallelization for MIO primal heuristics remains an open challenge.

In this paper, we develop such a GPU-accelerated primal heuristic. Our algorithm, Parallel Node Generation and Exploration Algorithm (PaNGEA), implements an efficient local-search procedure on restricted subspaces (or nodes). Instead of relying on a single heuristic rule to construct these nodes, we leverage GPU capabilities to generate and evaluate multiple candidate nodes in a batched way, and demonstrate its performance on a testbed of 283 instances.

\subsection{Problem Setting}
We consider a general mixed-integer linear optimization problem
\begin{equation}
\min_{\bm{x} \in \mathbb{R}^n} \ \bm{c}^\top \bm{x}
\quad
\text{s.t.} \quad
\bm{A}\bm{x} \le \bm{b},
\quad
\bm{\ell}^0 \le \bm{x} \le \bm{u}^0,
\quad
x_j \in \mathbb{Z}, \ \forall j \in \mathcal{I},
\tag{$\mathcal{P}$} \label{eqn:p}
\end{equation}
where $\bm{c} \in \mathbb{R}^n$, $\bm{A} \in \mathbb{R}^{m \times n}$, $\bm{b} \in \mathbb{R}^m$, $\bm{\ell}^0, \bm{u}^0\in \mathbb{R}^n$, and
$\mathcal{I} \subseteq [n]:=\{1,\dots,n\}$ is the index set of integer variables. It will be convenient to consider a restricted version of Problem \eqref{eqn:p}, also called a subproblem or node.

\begin{definition}[Node]
For any pair of vectors $\bm{\ell}, \bm{u} \in \mathbb{R}^n$ satisfying $\bm{\ell}^0\leq\bm{\ell}\leq\bm{u}\leq\bm{u}^0$, we call Problem \eqref{eqn:p} with the additional constraints $\bm{\ell}\leq \bm{x} \leq \bm{u}$
a \emph{node} and denote it $\mathcal{P}(\bm{\ell}, \bm{u})$.
\end{definition}
For concision, we denote by $[\bm{\ell},\bm{u}]$ the set of vectors $\bm{x}$ such that $\bm{\ell}\leq \bm{x} \leq \bm{u}$.
By definition, we have $\mathcal{P} = \mathcal{P}(\bm{\ell}^0, \bm{u}^0)$, which is commonly referred to as the root node. Further, if $\ell_j = u_j$ for some $j \in [n]$, then $\mathcal{P}(\bm{\ell}, \bm{u})$ fixes coordinate $j$ of $\bm x$ to the common value $\ell_j = u_j$. We call the (linear) \emph{relaxation} of $\mathcal{P}(\bm{\ell}, \bm{u})$ the optimization problem obtained by dropping the integrality constraints.

\subsection{Related Work}
The current algorithmic framework of MIO solvers involves multiple components, including continuous relaxations (lower bounds), branching, node selection, and feasible solution heuristics (upper bounds).

With the GPU acceleration of continuous optimization algorithms, such as the primal-dual linear programming (PDLP) method of \citet{lu2025cupdlp}, a natural direction is to use GPUs for solving the continuous relaxations of MIO problems faster. 
For instance, promising results have been obtained by using GPU-accelerated versions of PDLP \citep{blin2026batched,de2026power}, the alternating direction method of multipliers \citep{meng2026gpu}, or proximal gradient algorithms \citep{liu2026sequential,liu2026gpu}. 

Several works investigate the deeper integration of GPU acceleration within a branch-and-bound scheme. 
\citet{blin2026batched,meng2026gpu,liu2026gpu} consider solving similar relaxations (corresponding to different active nodes in the branch-and-bound tree) in a batched way. 
\citet{liu2026sequential} transfer additional node processing steps to the GPU, including rounding heuristics and branching selection. 
Beyond MIO, \citet{zhang2025accelerating} leverage GPU parallelization in a spatial branch-and-bound algorithm for global (non-convex) optimization.

Another central component is finding high-quality feasible solutions, which typically relies on primal heuristics.
A major idea in primal heuristics is to search for a high-quality solution within a restricted subspace or neighborhood.
Classic examples include local branching~\citep{fischetti2003local} or relaxation-induced neighborhood search~\citep{danna2005exploring}. However, these techniques rely on solving a small MIO problem. 
Several alternatives have been proposed to circumvent the need for an MIO solver:
diving \citep{guzelsoy2013restrict} and fix-and-propagate~\citep{salvagnin2025fix} conduct a depth-first search on the branch-and-bound tree to fix variables iteratively; feasibility pump
\citep{fischetti2005feasibility} iterates between solving relaxations and projection; move-based approaches, such as feasibility jump \citep{luteberget2023feasibility} and Local-MIP \citep{lin2025local}, make one-coordinate changes on the decision variables to greedily construct better solutions.
These heuristics are potentially easier to accelerate on GPUs because they rely on solving linear relaxations or on performing a large number of simple algebraic operations. Examples include the GPU acceleration of fix-and-propagate \citep{kempke2025low}, local search \citep{ccorduk2025gpu, tjusila2026chap}, rounding, and bound propagation \citep{ccorduk2025gpu}. 

It is worth mentioning that GPU acceleration of several primal heuristics has been discussed for the cases of specific problem classes such as vehicle routing \citep{schulz2013efficient,abdelatti2020improved}, scheduling \citep{czapinski2011tabu}, or pure-binary optimization problems \citep{abbas2022fastdog,wei2025gfors}. For a broader survey, see \citet[][section~2]{neira2024utilizing}.

Our work contributes to the literature on general-purpose GPU-accelerated primal heuristics for mixed-integer linear optimization. As in \citet{ccorduk2025gpu} and \citet{tjusila2026chap}, which are the closest works to ours, our algorithm relies on an efficient GPU implementation of the Local-MIP procedure \citep{lin2025local}. 
Our algorithm has two distinctive features. 
First, instead of maintaining one search region, we leverage the batching capabilities of GPUs to conduct several local searches in parallel on different nodes. 
Second, we develop a two-stage implementation of Local-MIP that is more efficient at finding high-quality solutions on a restricted subproblem (in particular, when some of the coordinates are fixed). 
We develop PaNGEA as a standalone primal heuristic, which does not rely on any MIO solver. We envision that its architecture could be integrated into a GPU-native branch-and-bound solver in the future.

\subsection{Contributions}
Our contributions can be summarized as follows. 

First, we develop an efficient GPU-friendly heuristic for exploring a given subproblem (or node), based on solving its linear relaxation and then applying a Local-MIP procedure. Instead of applying Local-MIP to the restricted subproblem directly, we propose a two-stage implementation, i.e., we impose the additional node bounds only after obtaining a globally feasible solution. We find that our two-stage implementation is significantly more effective at finding feasible and high-quality solutions than the naive implementation of Local-MIP (reducing the average optimality gap by 30\%), and that warm-starting with the linear relaxation reduces the optimality gap by a factor of five. 

Second, we use GPU batching not only to accelerate node exploration, but to diversify how nodes are generated: instead of relying on a single node-generation heuristic like traditional solvers, PaNGEA evaluates multiple strategies in parallel and retains the most promising nodes at each iteration. GPU batching not only provides acceleration of traditional methods (our single-node GPU implementation of Local-MIP reduces the gap integral by 8\% and 18\% on the MIPcc26 and MIPLIB instances, respectively) but also enables parallel exploration of multiple node-generation rules (which reduces the gap integral further by 19\% and 12\%, respectively).

The rest of the paper is organized as follows: We present our overall algorithm and its different components in Section \ref{sec:alg}, and evaluate it numerically in Section \ref{sec:exp}. We discuss the opportunity to integrate our algorithm into a GPU-native branch-and-bound in Section \ref{sec:bnb}.

\section{Algorithm} \label{sec:alg}
Our algorithm relies on a new efficient strategy for conducting local search over a subproblem (Section~\ref{sec:alg.subroutine}). In particular, this strategy can be effectively conducted on GPUs in a batched way. We leverage this design in our main algorithm (Section~\ref{sec:alg.overview}), which iteratively generates and explores different subproblems in parallel (Section~\ref{sec:alg.node.generation.strategies}) and retains the most promising ones at each iteration (Section~\ref{sec:alg.node.selection}). We discuss the details of our GPU implementation and batching in Section~\ref{sec:alg.gpu}.

\subsection{Core Subroutine} \label{sec:alg.subroutine}
At the core of our algorithm is a two-stage implementation of the local-search procedure of \citet{lin2025local}, tailored to improve its performance when restricted to subproblems. Our improvements rely on two ingredients. First, we initialize the search (including the one at the root node) at the solution of the continuous (linear) node relaxation, instead of an all-zero vector as in \citet{lin2025local}. Second, we delay imposing the node bounds $\bm{\ell} \leq \bm{x} \leq \bm{u}$ until after a feasible solution has been found.

For the root node $\mathcal{P}(\bm{\ell}^0,\bm{u}^0)$, Local-MIP \citep{lin2025local} starts from an initial point $\bm{x}^{0}$ (an all-zero vector in their implementation), rounds the integer coordinates of $\bm{x}^{0}$ to the nearest integer,  projects it onto the box $[\bm{\ell}^0,\bm{u}^0]$, and iteratively applies one-coordinate changes of the form ${x}_j^{t+1}={x}_j^{t} + \delta^t_j$. \citet{lin2025local} describe several rules to construct admissible updates $\delta^t_j$ as well as a scoring system used to pick which coordinate $j \in [n]$ to update. Typically, the scoring system favors moves that reduce constraint violation (when the incumbent is infeasible) or improve the objective value.  Variable bounds are treated differently than other constraints: the step sizes $\delta_j^t$ are restricted to $[\ell_j^0 - x_j^t, u^0_j - x_j^t]$. Because the initial point after rounding and projection belongs to $[\bm{\ell}^0,\bm{u}^0]$, we always have $\bm{x}^t \in [\bm{\ell}^0,\bm{u}^0]$.

For a given node $\mathcal{P}(\bm{\ell}, \bm{u})$, the naive way of doing local search is to apply the procedure described above with the tighter bounds $(\bm{\ell}, \bm{u})$. Because local search makes myopic one-coordinate changes, it sometimes needs to break feasibility and violate constraints in order to find feasible or better solutions in the end. If so, restricting the variable bounds (and, in many cases, fixing some of the coordinates), hence limiting the range of admissible step sizes,  may negatively impact performance. 
Instead, we make two algorithmic changes to \emph{incentivize} solutions within $[\bm{\ell}, \bm{u}]$.

First, we run a local-search procedure on the entire space, i.e., for $\mathcal{P}(\bm{\ell}^0, \bm{u}^0)$, but initialized with the solution of the linear relaxation of $\mathcal{P}(\bm{\ell}, \bm{u})$. In this case, $\bm{x}^0 \in [\bm{\ell}, \bm{u}] \subseteq [\bm{\ell}^0,\bm{u}^0]$. 
At each iteration, the admissible step sizes ensure that iterates remain in $[\bm{\ell}^0, \bm{u}^0]$, but they can leave the subregion $[\bm{\ell}, \bm{u}]$.
Second, once a feasible solution to \eqref{eqn:p} is found, we impose the node variable bounds $(\bm{\ell}, \bm{u})$ by restricting the step sizes to $\delta_j^t \in [\ell_j - x_j^t, u_j - x_j^t]$. This implementation does not immediately project the next iterates onto $[\bm{\ell},\bm{u}]$ but rather incentivizes convergence towards a solution that belongs to the node. 

\subsection{Overview of PaNGEA} \label{sec:alg.overview}
We briefly describe our algorithm in this section, which is visually summarized in Figure~\ref{app:fig:full.diagram}. We provide additional implementation details in Appendix~\ref{sec:app.algorithm}.
\begin{figure}
\FIGURE
{\includegraphics[width=.9\textwidth]{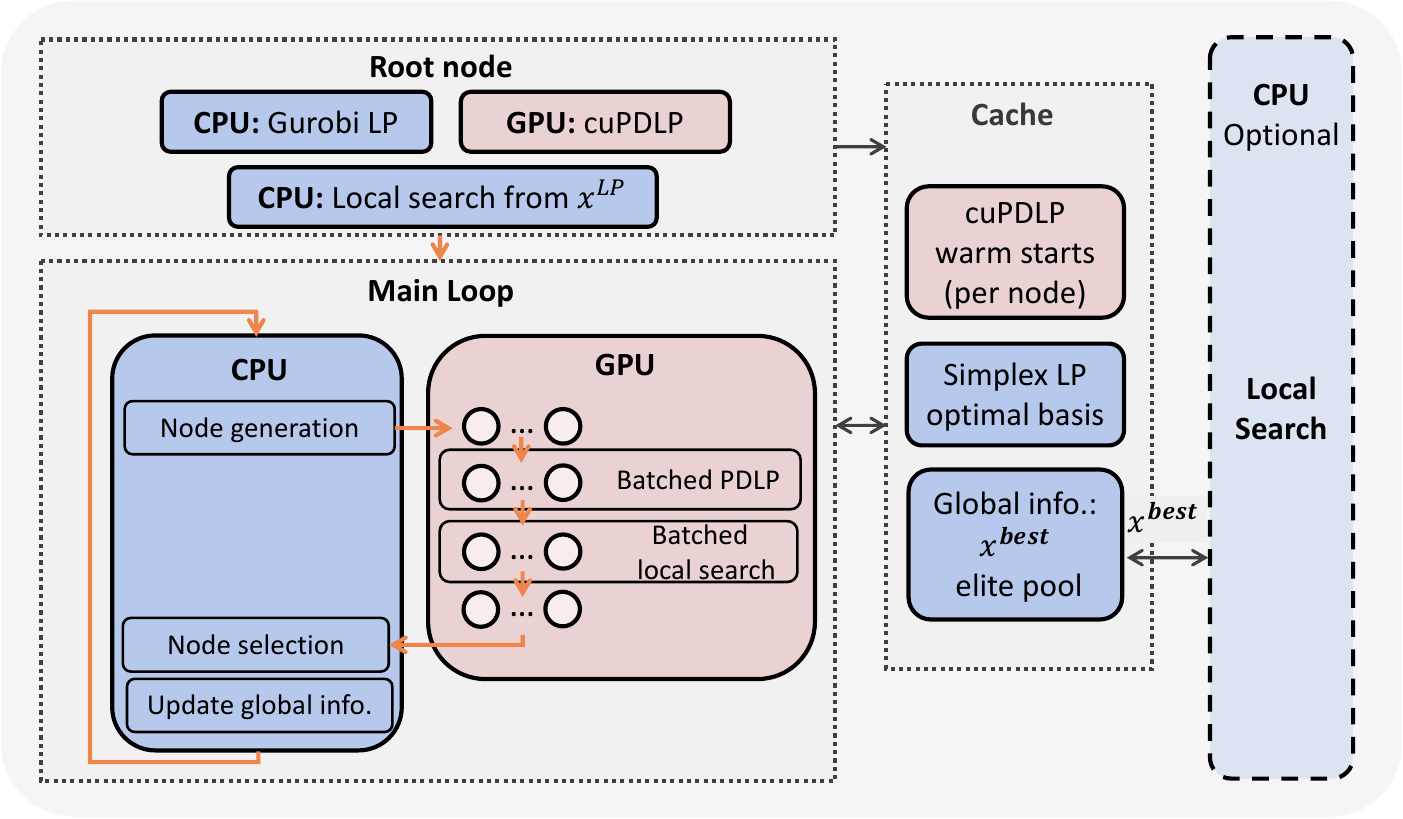}}
{{Visual representation of the architecture of PaNGEA. }\label{app:fig:full.diagram}}
{}
\end{figure}

First, we apply our core subroutine (linear relaxation followed by local search) from Section~\ref{sec:alg.subroutine} at the root node. We duplicate the root node into $W$ nodes, $\mathcal{N}_1,\dots,\mathcal{N}_W$.

The main part of our algorithm (the main loop) then follows. From each node $\mathcal{N}_w, w \in [W]$ we construct $R$ new nodes, $\mathcal{N}_{w,1},\dots,\mathcal{N}_{w,R}$ by applying different node-generation strategies (see Section~\ref{sec:alg.node.generation.strategies}). We leverage the batching capabilities of GPUs to apply the core subroutine to all $W \times R$ nodes $\{\mathcal{N}_{w,r}\}$ in parallel. We comment on the GPU implementation of the subroutine in Section~\ref{sec:alg.gpu}. We then retain only the best $W$ nodes and proceed to the next iteration (Section~\ref{sec:alg.node.selection}).

We keep in memory (on the CPU) the best solution found $\bm{x}^{\texttt{best}}$, its objective value $\texttt{bestUB}$, and a pool of high-quality solutions $\mathcal{E}$ ($|\mathcal{E}|=20$ in our implementation). In addition, we use a separate CPU thread for local search. Every time the best incumbent $\bm{x}^{\texttt{best}}$ is updated by the main loop, we restart the local search from $\bm{x}^{\texttt{best}}$ on this CPU thread. The separate CPU thread serves as a more focused local search while broader exploration of the feasible space is performed on the GPU. Such a hybrid CPU--GPU implementation is common in recent developments of (discrete) optimization algorithms, to leverage the strengths of each hardware platform and provide more robust performance \citep[see, e.g.,][]{tjusila2026chap}; we discuss the benefit of having both a GPU and a CPU local search in Section \ref{sec:exp:cpugpu}.

\subsection{Node-Generation Strategies} \label{sec:alg.node.generation.strategies}
We now describe six strategies for generating alternative nodes to explore. Unlike traditional CPU-based approaches, our algorithm does not rely on picking one strategy but instead applies all of them in parallel at each iteration.

The first two strategies modify the bounds of a parent node $\mathcal{P}(\bm{\ell}, \bm{u})$, for which we have access to primal and dual solutions of the linear relaxation, $\bm{x}^{\mathtt{LP}}$ and $\bm{y}^{\mathtt{LP}}$, as well as a solution from the local-search procedure, $\bm{x}^{\mathtt{LS}}$.

\paragraph{Reduced costs.} From the linear relaxation, we can compute the reduced costs $\bm{r} = \bm{c} - \bm{A}^\top \bm{y}^{\mathtt{LP}}$. Our local search starts from $\bm{x}^{\mathtt{LP}}$ and returns $\bm{x}^{\mathtt{LS}}$. We can thus assign a cost $s_j = r_j \bigl(x^{\mathtt{LS}}_j - x^{\mathtt{LP}}_j\bigr)$ to each coordinate change. Let $\mathcal{J} \subseteq \mathcal{I}$ denote the indices with the lowest cost $s_j$ (typically, we fix $| \mathcal{J}| / |\mathcal{I}| = 4\%$ in our experiments). These correspond to the local-search moves with the lowest compromise in objective value.
For each $j \in \mathcal{J}$, if $\ell_j < u_j$ (unfixed coordinate), we set $\ell_j = u_j = x_j^{\mathtt{LS}}$ (we fix it), otherwise we set $\ell_j = \ell_j^0$ and $u_j = u_j^0$.
This fixing/unfixing strategy is similar in spirit to restrict-and-relax search \citep{guzelsoy2013restrict}.

\paragraph{Diving.}
Starting from the current bounds $(\bm{\ell}, \bm{u})$, we randomly select $j \in \mathcal{I}$ such that $\ell_j < u_j$. If $x_j^{\mathtt{LP}} \in \mathbb{Z}$, we set $\ell_j = u_j = x_j^{\mathtt{LP}}$. Otherwise, we solve two linear relaxations, with the additional constraint $x_j = \lfloor x_j^{\mathtt{LP}} \rfloor$ or $x_j = \lceil x_j^{\mathtt{LP}} \rceil$, respectively, and fix $\ell_j = u_j$ to the value with the tighter (i.e., larger) relaxation bound. We repeat this process until a target number of variables has been fixed.

\vspace{6pt}
The next two strategies rely on the solution pool maintained on the CPU, $\mathcal{E} = \{ \bm{x}^{(e)}, e=1,\dots,E\}$, and are described for the subset of indices $\mathcal{I}_b \subseteq \mathcal{I}$ that corresponds to binary variables only.

\paragraph{Consensus restart.} Starting from $(\bm{\ell},\bm{u}) = (\bm{\ell}^0,\bm{u}^0)$ we fix $\ell_j=u_j$ to 0 (resp. 1) if at least 80\% of the solutions in the pool $\mathcal{E}$ satisfy $x_j^{(e)} = 0$ (resp. 1).

\paragraph{Elite.} Starting from the binary coordinates that are currently fixed, the Elite strategy first unfixes and then fixes some of these coordinates. These decisions are made by following the most common choice in the elite pool. We refer to Appendix~\ref{sec:app.algorithm} for a more formal description of the criteria used.

\vspace{6pt}
We consider one strategy aimed at exploring the neighborhood of 
$\bm{x}^{\texttt{best}}$.
\paragraph{Best.} Starting from $(\bm{\ell},\bm{u}) = (\bm{\ell}^0,\bm{u}^0)$, for each $j \in \mathcal{I}$, we set $\ell_j = u_j = {x}_j^{\texttt{best}}$ with probability $90\%$, independently.

\vspace{6pt}
Finally, we consider one strategy with bound propagation to address instances where finding feasible solutions is challenging. We only activate this strategy on instances where the root node analysis provided no feasible solution.

\paragraph{Deep diving with bound propagation.} Starting from $(\bm{\ell},\bm{u}) = (\bm{\ell}^0,\bm{u}^0)$, we sample $j \in \mathcal{I}_b$, set $\ell_j = u_j = 1$, and apply bound propagation. We repeat this process until all binary variables are fixed or a conflict in bound propagation is raised.

\subsection{Node Selection}
\label{sec:alg.node.selection}
After exploring all nodes $\{\mathcal{N}_{w,r}\}$ in one iteration of the main loop, we retain the `best' $W$ nodes for the next iteration. Our selection is based on the following criteria:
\begin{itemize}
    \item We discard suboptimal nodes, i.e., such that $\bm{c}^\top \bm{x}^{\texttt{LP}} > \texttt{bestUB}$ when $\bm{x}^{\texttt{LP}}$ is optimal.
    \item We sort nodes by increasing objective value, $\bm{c}^\top \bm{x}^{\texttt{LS}}$, and select the best $W$ nodes.
\end{itemize}
If a node relaxation is infeasible (`$\bm{c}^\top \bm{x}^{\texttt{LP}} = +\infty$'), the first criterion discards the node unless no feasible incumbent has been found yet (i.e., $\texttt{bestUB} = +\infty$). Furthermore, if this criterion leaves fewer than $W$ nodes to choose from, we randomly decide to keep some of the discarded nodes and unfix 10\% of the coordinates they currently fix.

For the second criterion, the top-$W$ nodes may include nodes for which the local search failed to find any feasible solution (`$\bm{c}^\top \bm{x}^{\texttt{LS}} = +\infty$'). In these cases, we order nodes by increasing number of violated constraints, $\sum_{j \in [m]} \mathbf{1}\left( \bm{a}_j^\top \bm{x}^{\texttt{LS}} - b_j > 0\right)$.

\subsection{GPU Implementation and Batching}
\label{sec:alg.gpu}
A central design of our algorithm is that each step of our core subroutine (node relaxation and local search) can be implemented on GPUs and that nodes only differ in their bounds. Therefore, we process multiple nodes in a batched fashion and benefit from GPU acceleration.
The shared static data is stored on the GPU once at the beginning of the algorithm, while node-specific data is stacked across the batch dimension and updated in parallel.

For solving the node relaxation, we use cuPDLP \citep{lu2025cupdlp}, a GPU implementation of a primal-dual hybrid gradient method for linear programming \citep[PDLP; see][]{applegate2023faster}.
PDLP solves the corresponding primal-dual saddle-point problem with first-order updates.
The dominant computations are sparse matrix-vector products $\bm{A}\bm{x}$ and coordinate-wise projections, both of which are GPU-friendly.
To process $B = W \times R$ nodes in parallel, we store the iterates $\bm x$ and node bounds $(\bm{\ell},\bm{u})$ as three matrices in $\mathbb{R}^{n \times B}$.
With this layout, each PDLP update is carried out for all nodes in parallel, using sparse matrix-matrix products and coordinate-wise projections. The batched formulation preserves the regular structure of PDLP while spreading the cost of GPU kernel launches across multiple nodes and further improving kernel throughput.

In the local-search algorithm, the dominant computations occur in generating and evaluating the candidate moves, $\delta_j^t$, which are all independent and thus parallelizable.
However, the GPU implementation is not straightforward, as there are multiple possible mappings between threads and computations, which may affect load balance. For example, a move $(j, \delta_j)$ may be computationally heavy to evaluate if the $j$-th column of $\bm{A}$ is dense (i.e., $x_j$ is involved in many constraints). In our implementation, we map a warp to $32$ nonzeros in $A_{\cdot j}$ for any $j$. We do not see a significant speedup from using CUDA graphs; see \citet{ccorduk2025gpu} for a discussion on a CUDA-graph implementation of Local-MIP.
As for solving the relaxations, all nodes in a batch share the same problem data, while node-specific quantities such as variable bounds and the search state are stored column-wise across the batch. The searches proceed independently for every node but use common GPU kernels to generate or evaluate candidate moves.

\section{Numerical Results} \label{sec:exp}
In this section, we first validate the effectiveness of our node exploration subroutine (Section~\ref{sec:exp:ls-subspace}). We demonstrate the benefit of using multiple ($R > 1$) node-generation strategies concurrently in Section~\ref{sec:exp:r} and the benefit of exploring multiple nodes in parallel ($W > 1$) in Section~\ref{sec:exp:w}. We then benchmark our method against alternatives in Section~\ref{sec:exp:benchmark}. In Section \ref{sec:exp:cpugpu}, we explore the benefits of having a CPU local search running in parallel to our GPU-batched search.

Throughout this section, we use the MIPcc26 testbed\endnote{\url{https://www.mixedinteger.org/2026/competition/}} and the MIPLIB2017 benchmarking set \citep{gleixner2021miplib}. We measure performance in terms of final (primal) gap at termination (which measures the quality of the best solution found) and (primal) gap integral (which captures the time/quality trade-off). {The `gap' is relative to a best-known objective, which in our case is the best objective found across all algorithms in the comparison as well as the `best solution' provided with the MIPcc26 and MIPLIB instances. Following \citet{berthold2013measuring}, the gap integral is defined as the integral of the relative gap (between 0 and 1) over time (in seconds). We use a 5-minute time limit so the gap integral ranges between 0 (the best-known solution is found immediately) and 300 (the gap stays at 1 throughout the 5-minute period).} We use an NVIDIA H100 (80GB) GPU for benchmarking (Section~\ref{sec:exp:benchmark}) and an NVIDIA A100 PCIe (80GB) in other experiments.

\subsection{Local-MIP with Node Pressure}
\label{sec:exp:ls-subspace}
We first demonstrate the effectiveness of our local-search subroutine (Section~\ref{sec:alg.subroutine}) on a subset of the MIPcc26 instances, namely instances $\{1,12,14,15,24,31,32,42,43,50\}$.
On each instance, we find a (near)-optimal solution using Gurobi and construct 21 subproblems by fixing some of the integral coordinates to their values in that solution.

\begin{figure}
\FIGURE
{\begin{subfigure}[b]{.44\hsize}
    \centering
    \includegraphics[width=\linewidth]{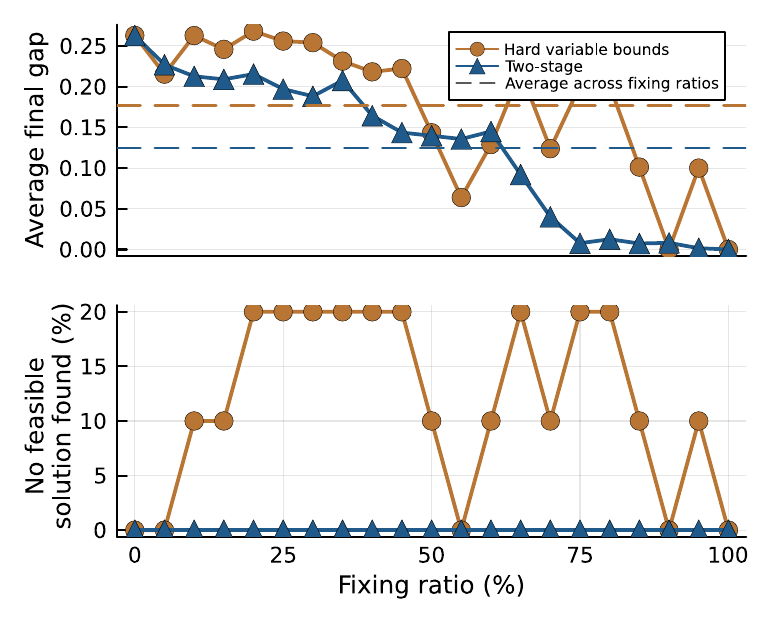}
    \caption{}\label{fig:ls-subspace.a}
    \end{subfigure}
    \begin{subfigure}[b]{.4\hsize}
    \centering
    \includegraphics[width=\linewidth]{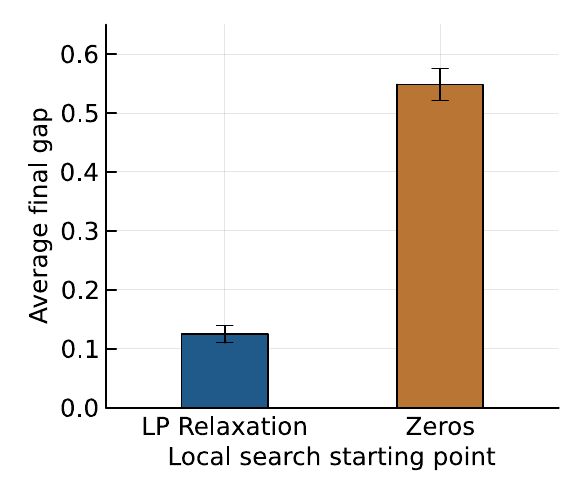}
    \caption{}\label{fig:ls-subspace.b}
    \end{subfigure}}
{Performance comparison between our two-stage local search and a naive implementation of \citet{lin2025local} on $\mathcal{P}(\bm{\ell},\bm{u})$. \label{fig:ls-subspace}}
{Panel (a) reports the average gap (top) and fraction of subproblems where the search fails to find a globally feasible solution (bottom), for both methods, averaged over instances with the same ratio of fixed integer coordinates. Panel (b) reports the average gap achieved by our two-stage local search when initialized with the node relaxation solution versus a vector of all zeros.}
\end{figure}

Figure~\ref{fig:ls-subspace.a} compares our two-stage procedure with the direct application of local search on the subproblem $\mathcal{P}(\bm{\ell},\bm{u})$ (`hard variable bounds'). Both methods are initialized with the solution of the node relaxation. In terms of average gap (top panel), both methods improve as we fix more coordinates (as expected). For a fixed fraction of fixed coordinates, however, our two-stage algorithm achieves significantly smaller gaps, with a reduction of five percentage points on average across different fractions of fixed coordinates. This improvement is partly due to the fact that the naive implementation of local search often fails even to find a feasible solution to \eqref{eqn:p} (bottom panel), even when a large fraction (70--90\%) of the coordinates are fixed, unlike our two-stage implementation.

We also evaluate the effectiveness of our initialization strategy in Figure~\ref{fig:ls-subspace.b}. Starting from the node-relaxation solution in our two-stage procedure leads to a substantially smaller average gap ($12\%$) than starting from zeros ($55\%$).
While \citet{lin2025local} describe their method as being `LP-free', our results confirm that it can benefit from being initialized with the linear relaxation.

\subsection{Diversifying Node-Generation Strategies} \label{sec:exp:r}
\begin{figure}
\FIGURE
{\centering
\begin{subfigure}[b]{.48\textwidth}
\centering
\includegraphics[width=\linewidth]{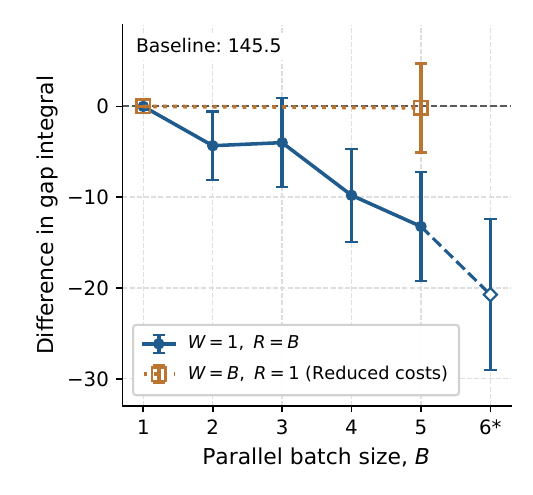}
\caption{Performance of diversifying strategies.}\label{fig:W=1.a}
\end{subfigure}\hfill
\begin{subfigure}[b]{.48\textwidth}
\centering
\includegraphics[width=\linewidth]{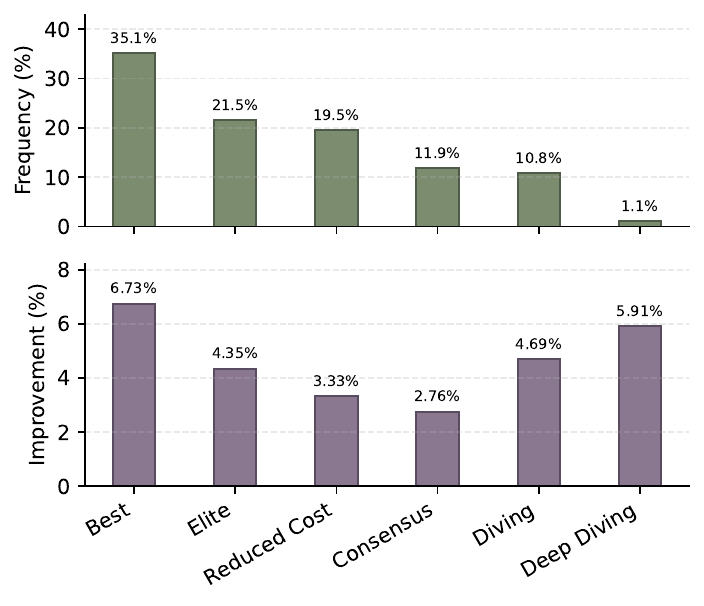}
\caption{Profile of node-generation strategies.}\label{fig:W=1.b}
\end{subfigure}}
{Parallel exploration and the contribution of node-generation strategies.\label{fig:W=1}}
{Panel (a) shows mean paired differences in gap integral over the MIPcc26 instances, relative to $W=R=B=1$ with Reduced Costs (baseline 145.5). Error bars are $\pm1$ across-instance standard error of the differences. The $R>1$ case uses a greedy sequence to decide the strategies for each $R$; see Appendix~\ref{app:strategy.parallelism}. The $R=6^*$ algorithm adds Deep Diving to $R=5$, which is only applied when the root local search does not find a feasible solution.
Panel (b) reports the profile of node-generation strategies for PaNGEA: the frequency at which a strategy finds the new best solution (top) and the relative improvement in \texttt{bestUB} (bottom).}
\end{figure}

We now evaluate the benefit of considering multiple node-generation strategies in parallel for our PaNGEA algorithm using the MIPcc26 instances.

Figure~\ref{fig:W=1.a} reports the average gap integral achieved by PaNGEA as the number of strategies $R$ increases, with $W=1$ node (so at each iteration, we consider $W=1$ node, generate and explore $W \times R = R$ new subproblems using the different node-generation strategies, and then retain the best node for the next iteration). We add the node-generation strategies progressively as described in Appendix~\ref{app:strategy.parallelism}. 

We observe that performance improves as $R$ increases, with $R=6$ achieving a gap integral 14\% lower than $R=1$. To emphasize the fact that the gains come from the different node-generation strategies and not simply from exploring multiple subproblems in parallel, we also report the performance of PaNGEA with $R=1$ and $W=5$ in Figure~\ref{fig:W=1.a}. Unlike $W=1,R=5$, PaNGEA with $W=5,R=1$ demonstrates no significant improvement over $W=1,R=1$.

To further understand these gains, whenever the best incumbent solution $\bm{x}^{\texttt{best}}$ is updated by the GPU search, we record which node-generation strategy generated this incumbent (which we refer to as a `success') and by how much \texttt{bestUB} is improved.
Figure~\ref{fig:W=1.b} reports the frequency and objective improvement associated with each rule. We observe that some strategies, such as Deep Diving, are rarely beneficial ($1.1\%$ of the time) but provide large improvements ($5.91\%$) when they do.
Others, such as Best or Elite, frequently generate new best incumbents ($35.1\%$ and $21.5\%$ of the time, respectively) and yield sizeable improvements ($6.73\%$ and $4.35\%$, respectively).

\FloatBarrier
\subsection{Parallel Exploration of Multiple Nodes} \label{sec:exp:w}
The number of nodes explored in parallel, $B = W \times R$, is a key parameter in PaNGEA. Increasing $W$ (hence, $B$) increases the amount of parallel work and typically improves GPU utilization and performance.
However, each main loop iteration terminates once all nodes $\{\mathcal{N}_{w,r}\}$ have completed their work. Hence, running more nodes in parallel raises the risk of one node slowing down the others. We focus on algorithm performance (gap integral, here) in this section but report results on GPU utilization in Appendix~\ref{app:gpu.utilization}.

\begin{figure}
\FIGURE
{\includegraphics[width=.45\textwidth]{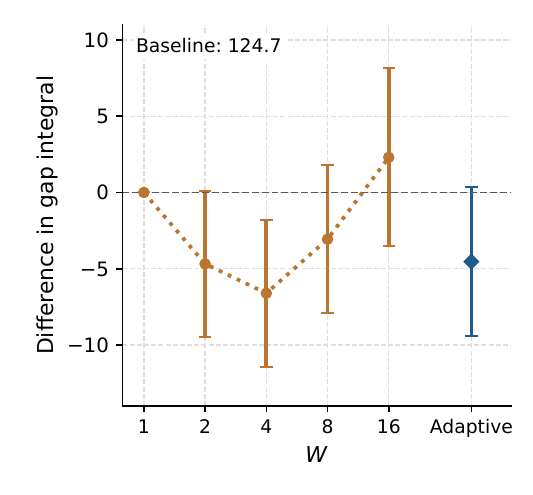}}
{Parallel Exploration of Multiple Nodes. \label{fig:increasing-W}}
{Mean paired differences in gap integral over MIPcc26 instances, relative to PaNGEA with $W=1, R=6$ (baseline 124.7). Error bars are $\pm1$ across-instance standard error. }
\end{figure}

Figure~\ref{fig:increasing-W} illustrates this trade-off by comparing the performance of PaNGEA for different values of $W$, with $R=6$ fixed. 
We observe that the average gap integral exhibits a U-shape.
For small values of $W$, the gap integral first decreases as $W$ increases: the average gap integral decreases from 124.7 at $W=1$ to 120.1 at $W=2$ and 118.1 at $W=4$. These results contrast with the nearly unchanged performance of the $R=1$ case in Figure~\ref{fig:W=1.a}, suggesting that additional parallel work is more useful when a range of node-generation strategies are used to generate distinct search directions.
For higher values of $W$, however, the gap integral starts to increase; for $W=16$, it even exceeds that of the $W=1$ baseline. Increasing $W$ therefore requires balancing broader exploration against the effort spent processing each batch. 

To address this trade-off, we propose an adaptive rule where $W$ is set according to the instance size (in bytes) and the available GPU memory. Precisely, we set $W$ to the closest positive integer to $\min(5 \cdot 10^{-3} \cdot \frac{\text{GPU memory}}{R \times \text{size of } \mathcal{P}},16)$.
We observe that our adaptive rule attains a gap integral of 120.2, substantially below the $W=1$ baseline and close to the best $W$ chosen in hindsight.

\subsection{Benchmarking} \label{sec:exp:benchmark}
We report the aggregated benchmarking results on the 50 MIPcc26
instances and 233 MIPLIB instances\endnote{We exclude seven instances that are proved infeasible.} in Tables~\ref{tab:agg.benchmarking} and~\ref{tab:agg.benchmarking.miplib}, respectively. All instances are processed through Gurobi presolve. We compare our algorithm PaNGEA (with $W$ adaptive and $R=6$) with a variant of PaNGEA without the CPU local-search thread, a non-parallel version of PaNGEA with $W=1$ and $R=1$, and a CPU implementation of Local-MIP\endnote{For Local-MIP, we follow \url{https://github.com/shaowei-cai-group/Local-MIP}. We do not compare with \citet{ccorduk2025gpu} and \citet{tjusila2026chap} because the corresponding benchmarking code was not available at the time of our experiment.} \citep{lin2025local}, Gurobi heuristics, and Gurobi heuristics without branching\endnote{We configure the ``heuristics'' mode of Gurobi by setting $\texttt{MIPFocus}=1,\texttt{Heuristics}=1$ and the ``without branching'' mode by setting $\texttt{Cuts}=3,\texttt{NodeLimit}=1,\texttt{MIPFocus}=1,\texttt{Heuristics}=1$.}. 
We report instance-level feasible objective value results in Tables~\ref{app:tab:computational.results.}--\ref{app:tab:computational.results.miplib.5}.
 
\begin{table}
\TABLE
{{MIPcc26 benchmarking results in terms of gap integral, final gap, and number of feasible instances.}\label{tab:agg.benchmarking}}
{\setlength{\tabcolsep}{6pt}
\begin{tabular}{@{}lrrr@{}}
\toprule
\multicolumn{1}{c}{Method} & Gap Integral & Final Gap & \# Feasible \\
\midrule
PaNGEA & \bf 112.94 & \bf 0.27 & \bf 44 \\
PaNGEA without CPU thread & 160.47 & 0.49 & 39 \\
PaNGEA W=1,R=1 & 138.87 & 0.40 & 41 \\
Local-MIP & 151.25 & 0.45 & 42 \\
\midrule
Gurobi heuristics & \bf 85.42 & \bf 0.20 & \bf 45 \\
Gurobi heuristics without branching & 111.86 & 0.30 & 45 \\
\bottomrule
\end{tabular}
}
{}
\end{table}

\begin{table}
\TABLE
{{MIPLIB benchmarking results in terms of gap integral, final gap, and number of feasible instances.}\label{tab:agg.benchmarking.miplib}}
{\setlength{\tabcolsep}{6pt}
\begin{tabular}{@{}lrrr@{}}
\toprule
\multicolumn{1}{c}{Method} & Gap Integral & Final Gap & \# Feasible \\
\midrule
PaNGEA & \bf 91.75 & \bf 0.24 & \bf 211 \\
PaNGEA without CPU thread & 112.90 & 0.33 & 203 \\
PaNGEA W=1,R=1 & 104.22 & 0.29 & 205 \\
Local-MIP & 126.83 & 0.39 & 203 \\
\midrule
Gurobi heuristics & \bf 29.20 & \bf 0.05 & \bf 226 \\
Gurobi heuristics without branching & 63.14 & 0.17 & 214 \\
\bottomrule
\end{tabular}
}
{}
\end{table}

\FloatBarrier

PaNGEA achieves the best performance among the non-commercial methods on both benchmark sets and on all three performance criteria (gap integral, final gap, number of feasible instances). It achieves average gap integrals of 112.94 on MIPcc26 and 91.75 on MIPLIB, and average final gaps of 0.27 on MIPcc26 and 0.24 on MIPLIB. Compared to the non-commercial baselines, PaNGEA achieves 12--30\% lower gap integral and 17--45\% lower final gap. In other words, it finds better feasible solutions faster on average. By looking at the distribution of final gap and gap integral over instances (see Figures~\ref{app:fig:benchmark.boxplot.integral}-\ref{app:fig:benchmark.boxplot.final.gap} in Appendix~\ref{app:benchmark.supplement}), we observe that 
PaNGEA not only improves the averages but shifts the distribution downwards, with noticeable benefit on the tail (i.e., the hardest instances). For example, it reduces the 75th percentile of the gap integral on MIPcc26 (resp. MIPLIB) from 268 (resp. 247) for Local-MIP down to 140 (resp. 146).

Comparing the performance of PaNGEA with that of PaNGEA without the CPU thread, we observe that the background CPU Local-MIP thread is an important component of our architecture, corroborating observations made for similar hybrid implementations in related contexts \citep[e.g.,][]{tjusila2026chap,liu2026sequential}. The gain is due to a different behavior of Local-MIP on CPUs vs. GPUs. On the GPU, batching increases the within-iteration workload at the expense of fewer iterations. Hence, running Local-MIP on the CPU thread can be seen as an intensification mechanism that explores the neighborhood of the best solution found so far more deeply. We further explore the contribution of the CPU thread to our performance in the following section.

We now compare the performance of PaNGEA with commercial methods. 
We note that Gurobi's performance is much stronger on the MIPLIB instances, most likely because these instances have been widely used to develop and engineer existing solvers.
On MIPcc26, PaNGEA achieves a smaller final gap than Gurobi without branching (0.27 vs 0.30), while their gap integrals are similar (112.94 vs 111.86). On MIPLIB, Gurobi without branching achieves a smaller gap integral and a smaller final gap (63.14 and 0.17 vs 91.75 and 0.24). Gurobi heuristics achieves a smaller final gap over both test sets. 

\begin{figure}
\FIGURE
{\includegraphics[width=.55\textwidth]{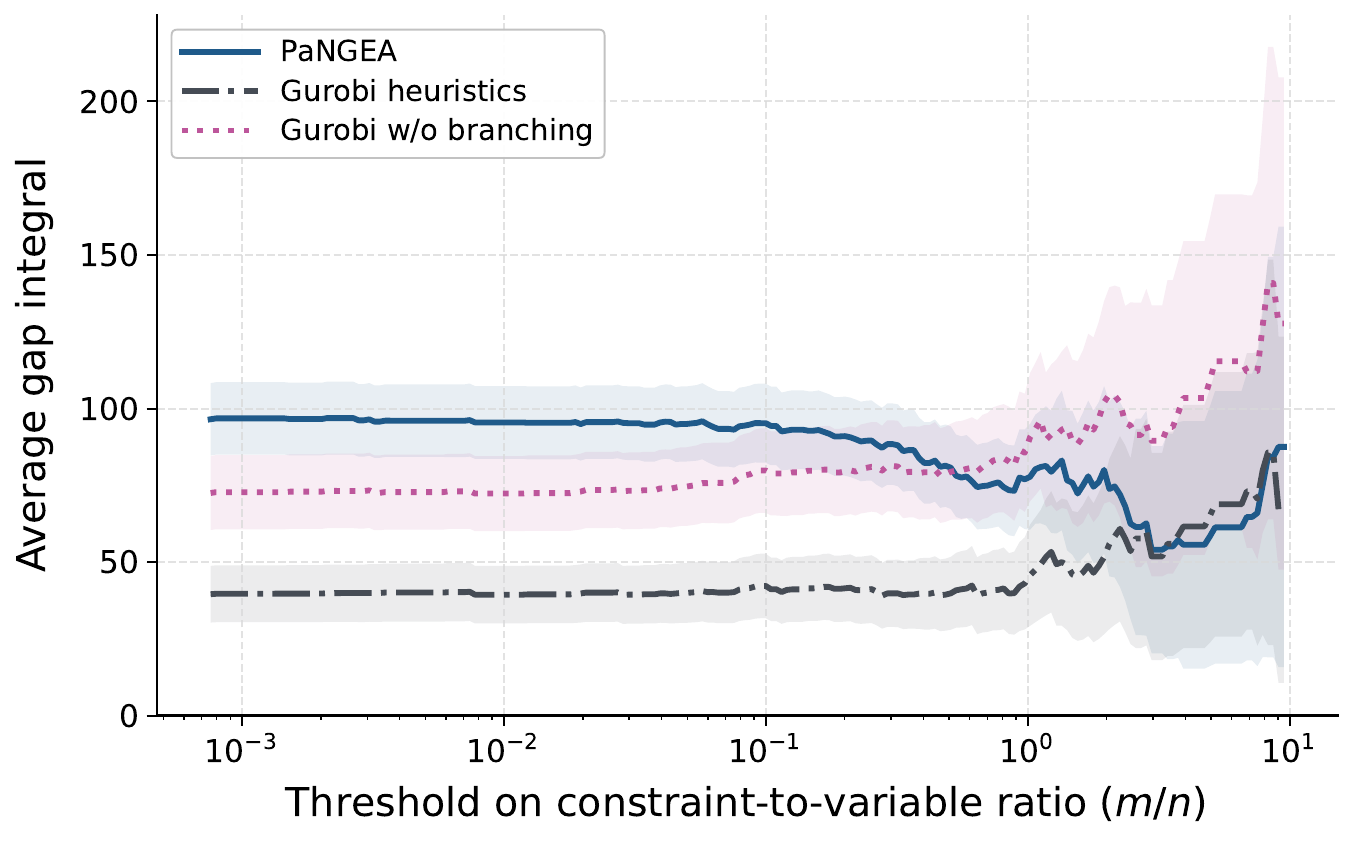}}
{{ Average gap integral over benchmark instances stratified by constraint-to-variable ratio.} \label{fig:benchmark.instance.size.integral}}
{The x-axis is a threshold $t$ on the constraint-to-variable ratio and is shown on a log scale. Each line represents the average gap integral over all instances with a constraint-to-variable ratio $m/n$ greater than $t$, as $t$ increases. Shaded regions are 95\% confidence intervals.}
\end{figure}

We find that a key driver in relative performance is the constraint-to-variable ratio, $m/n$. 
Figure~\ref{fig:benchmark.instance.size.integral} shows the average gap integral over instances such that $m/n$ is greater than a threshold (x-axis).
We observe that Gurobi's performance tends to deteriorate (gap integral increases) as the threshold $t$ on the constraint-to-variable ratio increases. In contrast, PaNGEA exhibits the opposite behavior. For example, PaNGEA achieves a lower average gap integral than Gurobi without branching on the instances with $m/n>0.5$, and than Gurobi heuristics on the instances with $m/n>3$. We observe a similar pattern for final gap as well (Figure~\ref{app:fig:benchmark.instance.size.final.gap}). This suggests that PaNGEA could serve as a useful complement to, rather than a substitute for, existing methods, especially for instances with a high constraint-to-variable ratio. We hypothesize that the poor performance of PaNGEA when $m/n$ is small could be due to the local-search procedure itself, which only changes one coordinate at each iteration (hence, is sensitive to high $n$ values), while updating the left-hand side of the $m$ constraints is efficiently handled on GPUs. 

\subsection{Value of a Hybrid GPU-CPU Algorithm}\label{sec:exp:cpugpu}

Finally, we further examine the complementary roles of our GPU node exploration and the parallel CPU local search. 

\begin{figure}
\FIGURE
{\centering
\includegraphics[width=0.85\linewidth]{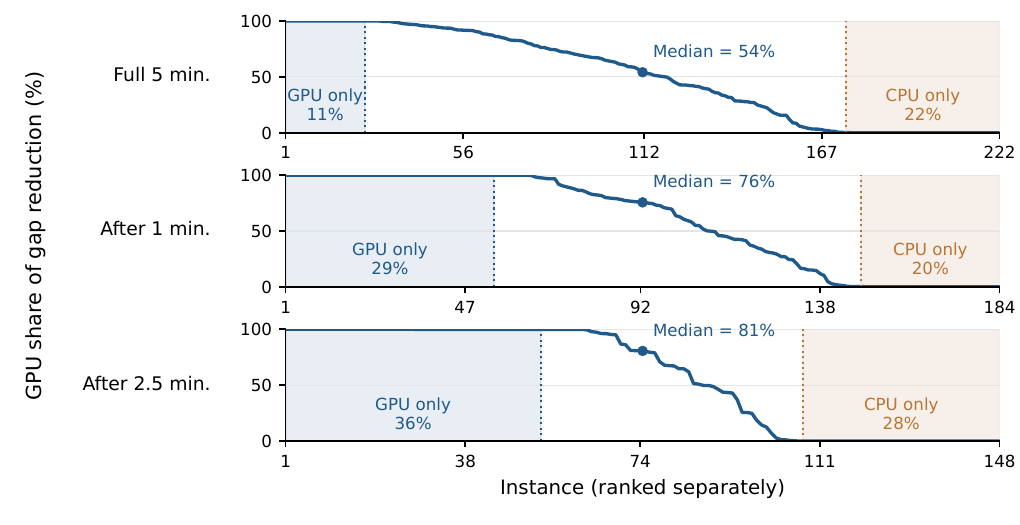}
}
{Instance-level GPU share of gap reduction over three time windows.\label{fig:benchmark.cpu.gpu.ranked.windows}}
{Every point on the curve represents, on one instance, the percentage of total gap reduction whose corresponding incumbent is produced by GPU batched local search. Instances (pooling MIPcc26 and MIPLIB) are ranked and indexed separately in each row. Shaded regions identify instances for which only one component contributes, i.e., GPU share is 1 or 0.}
\end{figure}

For each instance in our testbed (as in Section~\ref{sec:exp:benchmark}), we compute the total gap reduction between the initial gap and the final gap. For each new best solution found $\bm{x}^{\texttt{best}}$, we compute its contribution to this total gap reduction and we record whether it was found by the GPU or the CPU search. 
Figure~\ref{fig:benchmark.cpu.gpu.ranked.windows} reports the contribution of the GPU to this total gap decrease (`GPU share of gap reduction') across all instances (sorted by decreasing share). 

First, we observe that the two components are complementary: on 11\% of the instances, new best solutions come only from the GPU search, while the CPU local search does all the work on 22\% of the instances. Still, there are 67\% of the instances where both components contribute. The median GPU contribution is 54\%.

When restricting our attention to the later parts of the algorithm (i.e., looking at gap reduction after 1 or 2.5 minutes), we observe that the median GPU contribution increases to 76\% and 81\%, respectively. Note that the total number of instances decreases as we look at later progress, as we focus our attention on instances where the algorithm made further progress after 1 and 2.5 minutes, respectively. These observations suggest that, as the search progresses, batched exploration on the GPU plays an increasingly important role in escaping local minima and finding improvements. While this is a general trend, it does not mean that the first solutions $\bm{x}^{\texttt{best}}$ are systematically generated by the CPU local search and that the GPU-generated solutions are only useful at later times. Looking at instance-wise profiles, we see all patterns possible: CPU then GPU, GPU then CPU, and even a constant switch between GPU and CPU solutions (see Figure~\ref{fig:benchmark.cpu.gpu.pattern.examples}).

\section{Discussion: Towards a GPU-Native Branch-and-Bound}
\label{sec:bnb}
Although we develop PaNGEA as a standalone primal heuristic, its architecture closely resembles that of a branch-and-bound: for every node, we first solve a linear relaxation and then search for a feasible solution (Section~\ref{sec:alg.subroutine}), hence obtaining a lower and an upper bound; when deciding which nodes to retain at the next iteration (Section~\ref{sec:alg.node.selection}), we prune suboptimal nodes by comparing their lower bound with the best solution found. The main difference is that, in a branch-and-bound, nodes are generated in a hierarchical, tree-structured fashion, which guarantees that the algorithm eventually performs an exhaustive evaluation of all possible solutions. In contrast, the unstructured node-generation strategy of PaNGEA generates deep nodes (i.e., nodes with a large fraction of entries fixed) quickly, which helps find good feasible solutions.\endnote{On the other hand, the main disadvantage of not using a structured node-generation strategy is that pre-processing techniques, such as bound propagation, can become inefficient. Indeed, there is no clear inclusion relationship between a node and its children, so efforts spent on bound tightening, for example, are only useful for the current node. We mitigate this issue in PaNGEA by introducing the deep diving with bound propagation node-generation strategy, but an implementation of PaNGEA alongside a branch-and-bound could better leverage these techniques.}

Still, because PaNGEA applies to every node the same two procedures as a branch-and-bound (relaxation and feasibility heuristic), there is a natural opportunity to integrate them together. Namely, we can run a branch-and-bound algorithm and PaNGEA in parallel. Their node-generation and management processes would remain separate but, at each iteration, the branch-and-bound nodes and the PaNGEA nodes would be explored simultaneously in a batched way. This would allow batching more linear-relaxation solves together, which could improve computational efficiency. For branch-and-bound on CPU, there is typically only one node analyzed at each iteration, but recent work on GPU implementation of branch-and-bound considered analyzing multiple nodes simultaneously \citep{meng2026gpu,liu2026sequential}. We implement such an algorithm (see Appendix~\ref{app:bnb}) and investigate the computational benefits.

Table~\ref{tab:batched_vs_sequential_all50} reports linear relaxation and branch-and-bound related metrics on the MIPcc26 instances. We compare two possible implementations: one where we solve the linear relaxations of the branch-and-bound nodes and PaNGEA nodes simultaneously in the same batch ('Batched') and one where we solve the relaxations of the branch-and-bound nodes first and then the PaNGEA nodes (`Sequential'). In Sequential, each batch gets the same iteration limit and half the time limit of the `Batched' implementation. Regarding the linear relaxation, we observe that batching significantly increases the number of iterations per second by 63.5\%, hence improving computational efficiency. Similarly, the fraction of node relaxations solved without timing out increases from 9.8\% to 12.1\%. For the branch-and-bound part of the algorithms, better linear-relaxation solves translate into more pruning, with an increase in the number of pruned nodes of 15.6\%. These results suggest that the batched integration substantially improves computational efficiency and the branch-and-bound performance in terms of pruning. 

However, Table~\ref{tab:batched_vs_sequential_all50} also highlights a current bottleneck for designing generic branch-and-bound algorithms for MIO on GPUs: as a linear optimization solver, PDLP is currently not as accurate as the simplex or interior point methods, meaning many node relaxations (nearly 88\% in our experiments) are not solved to optimality nor proved infeasible. This, in turn, leads to weak bounds and slow progress. Current GPU implementations of branch-and-bound \citep[notably][]{meng2026gpu,liu2026sequential} instead focus on a narrower class of problems, for which alternative algorithms (such as coordinate descent) are extremely fast for solving the relaxed problems and can be efficiently implemented on GPUs. 

\begin{table}
\TABLE
{LP metrics and pruning outcomes under batched and sequential LP processing on the MIPcc26 instances.\label{tab:batched_vs_sequential_all50}}
{
\setlength{\tabcolsep}{5pt}
\begin{tabular}{@{}llrrr@{}}
\toprule
& Metric & Batched & Sequential & $\Delta$ (\%) \\
\midrule
\multirow{2}{*}{Node LP relaxation}
&  Node-iterations (PDLP steps) per second
& {3,376} & 2,065 & $+63.5$ \\
& Non-timeout rate
& {12.1\%} & 9.8\% & $+23.8$ \\
\midrule
Branch-and-bound
& Number of pruned nodes per instance
& 86.06 & 74.42 & $+15.6$ \\
\bottomrule
\end{tabular}
}
{Metrics are averages across instances. Node-iterations are the sum of effective PDLP steps over all nodes, excluding those on the nodes that are already optimal/infeasible. Non-timeout rate is the portion of branch-and-bound nodes whose linear relaxation is solved to optimality or proved infeasible under accuracy tolerance. $\Delta=100(\text{Batched}/\text{Sequential}-1)$.}
\end{table}

\section{Conclusion}
We introduce PaNGEA, a primal heuristic for MIO that uses GPUs not only for acceleration but also to explore multiple node-generation strategies in parallel. The key idea is to replace heuristic selection with batched heuristic exploration: rather than betting on one rule, PaNGEA evaluates several candidate node-generation strategies simultaneously and continues from the most promising nodes. Empirically, this design leads to the strongest performance among the non-commercial methods in the MIPcc26 and MIPLIB benchmarks, with clear gains over non-parallel variants. More broadly, our results suggest that GPU parallelism can enable new heuristic architectures for MIO, not just faster implementations of existing ones.

\begingroup \parindent 0pt \parskip 0.0ex \def\enotesize{\normalsize} \theendnotes \endgroup
\setcounter{endnote}{0}



\bibliographystyle{informs2014} 
\bibliography{reference} 

\newpage
\begin{APPENDICES}
\numberwithin{figure}{section} 
\numberwithin{table}{section} 
\numberwithin{algocf}{section} 
\numberwithin{proposition}{section} 
\numberwithin{corollary}{section} 
\numberwithin{remark}{section} 

\section{Detailed Algorithm Description} \label{sec:app.algorithm}
We now provide details on the implementation of each component in the PaNGEA algorithm (see Figure~\ref{app:fig:full.diagram}).

\subsection{Initialization}
The algorithm starts by analyzing the root node using our core subroutine.

First, we solve the root node relaxation using two solvers concurrently on different threads: a primal-dual first-order method (PDLP) and the commercial solver Gurobi. Solving the root node relaxation is important since its solution serves as a warm start for other relaxation solves. We use both solvers concurrently to achieve robust performance, since the performance of a specific method is highly instance-dependent.

Then, we run local search from the relaxed solution. If the local search fails to find a feasible solution, we consider the $R=6$ node-generation strategies from Section~\ref{sec:alg.node.generation.strategies}, including the \emph{deep diving with bound propagation} strategy. Otherwise, we only consider the other $R=5$ strategies.

\subsection{Implementation of the core subroutine}

\paragraph{Stopping criteria.}
Our main objective is to encourage a diverse exploration of the feasible space, so we are comfortable stopping each step of the process (node relaxation and local search) early. For the node relaxation, in addition to the PDLP convergence criteria, we impose a 5-second time limit. For the local search, we stop after 5 seconds, $2,000,000$ iterations, or if there is no improvement on the best solution found for more than 2 seconds.

\paragraph{Fallback.}
There are cases in which solving the node relaxation on GPUs with PDLP is less efficient than using a commercial solver like Gurobi on CPUs. To ensure our algorithm remains competitive, we implement the following strategy: when a node has at least $95\%$ of integral coordinates fixed and fewer than $10,000$ effective variables, or when the root node relaxation was solved (on CPU) in less than ${5}/{B}$ seconds, we use Gurobi to solve the node relaxation rather than batching it with the other node relaxations on the GPU.

Note that we use the output from the node relaxation $\bm{x}^\mathrm{LP}$, $\bm{y}^\mathrm{LP}$ as initialization for the local-search procedure. When using PDLP, we take the last iterate, even when the algorithm did not converge (because of the time limit or infeasibility). When using Gurobi, however, it may not provide a candidate $(\bm{x}^\mathrm{LP}, \bm{y}^\mathrm{LP})$ if the algorithm times out or if the relaxation is infeasible. To circumvent the issue, whenever we decide to solve at least one node relaxation with Gurobi instead of PDLP, we first run these Gurobi solves. If a node relaxation is either infeasible or times out with Gurobi, we then solve it again with PDLP on the GPU (batched with the other node relaxations), to obtain an initial $(\bm{x}^\mathrm{LP}, \bm{y}^\mathrm{LP})$ for the local search.

\paragraph{Partial solve.}
After each local-search procedure, we perform partial optimization on the best node (with the best feasible objective), i.e., we optimize the objective with respect to the continuous variables while all integer variables are fixed at their current value. This (highly) restricted linear optimization problem can be solved very efficiently (we use Gurobi on the CPU). To avoid delaying the local-search iteration too much, we impose a 0.5-second time limit on the solver.

\subsection{Caching}
We store the instance-specific data persistently on both the GPU and CPU, namely, $\bm{\ell}^0$, $\bm{u}^0$, $\bm{A}$, $\bm{b}$, $\bm{c}$, $\mathcal{I}$, and $\mathcal{I}_b$.

To efficiently warm-start the PDLP algorithm at each iteration of the main loop, we cache the terminal state of the PDLP algorithm for each node and use it to warm-start the algorithm for child nodes. We also have a {CPU-side} cache for the linear optimization solver (optimal basis). Recall that the CPU linear optimization solver is needed (i) to quickly solve node relaxations in the diving strategy; (ii) when doing partial solves in the local-search procedure; or (iii) when it is expected to be more efficient than PDLP on GPUs for node relaxations.

Finally, we keep a pool of `elite' solutions, $\mathcal{E}$. We limit the size of the elite pool to $|\mathcal{E}| \leq 20$. For memory efficiency, only the binary coordinates $\bm{x}_{\mathcal{I}_b}$ are stored.
Note that the elite pool does not necessarily contain feasible solutions. Before the first feasible solution is found, the elite pool consists of `promising' infeasible solutions (typically evaluated based on constraint violation). Once a feasible solution is found, all infeasible solutions are removed from the pool.

The elite pool is populated with objective value and diversity in mind.
When a new feasible solution is found, we add it to the pool if $|\mathcal{E}| < 20$. If $|\mathcal{E}| = 20$, we compute the Hamming distance between the new solution and all the solutions in the pool. If no elite solution is at a distance less than 5, the new solution is compared to the worst elite solution and replaces it provided it achieves a better (i.e., lower) objective value. If there exists an elite solution at a distance less than 5, the new solution is compared to the closest elite solution and replaces it provided it achieves a better (i.e., lower) objective value.

\subsection{Details about node-generation strategies}
\paragraph{Elite.} The \emph{Elite} node-generation strategy first unfixes and then fixes some of the binary coordinates in $\mathcal{I}_b$. Let $\mathcal{F}_0$ and $\mathcal{F}_1$ denote the sets of binary coordinates fixed to 0 and 1 in the current node, respectively.

For any $j \in \mathcal{I}_b$, we compute the average value of $x_j$ in the pool: $p_{\mathrm{base}}(j)=\frac{1}{E}\sum_{e=1}^E x^{(e)}_j$. For every $k \in \mathcal{F}_1$ (i.e., we are currently fixing $x_k = 1$), $p_{\mathrm{base}}(k)$ indicates the fraction of elite solutions that also have $x_k = 1$. Similarly for $\mathcal{F}_0$ and $1-p_{\mathrm{base}}(k)$. Accordingly, we sort $k \in \mathcal{F}_0 \cup \mathcal{F}_1$ according to the so-called alignment
\[
  a_\texttt{unfix}(k)=
  \begin{cases}
    p_{\mathrm{base}}(k) & \text{if } k \in \mathcal{F}_1,\\
    1-p_{\mathrm{base}}(k) & \text{if } k \in \mathcal{F}_0.
  \end{cases},
\]
and unfix coordinates with the lowest alignment. We update the sets
$\mathcal{F}_0$ and $\mathcal{F}_1$ accordingly.

Then, for each unfixed binary coordinate $j \in \mathcal{I}_b \setminus (\mathcal{F}_0 \cup \mathcal{F}_1)$, we compute a score $a_\texttt{fix}(j)$ to capture how popular $x_j=1$ or $x_j=0$ is among the elite pool. Formally, we set $a_\texttt{fix}(j) = \lambda\, p_{\mathrm{base}}(j) + (1-\lambda)\, {p}_{\mathrm{corr}}(j)$ with $\lambda = 1/2$ and
\begin{align*}
    {p}_{\mathrm{corr}}(j) = \dfrac{1}{|\mathcal{F}_1|}
       \sum_{k \in \mathcal{F}_1} \left( \dfrac{1}{|\mathcal{E} \cap \{ \bm{x} : x_k = 1 \}|} \sum_{e \in \mathcal{E} : x_k^{(e)} = 1} x^{(e)}_j\right).
\end{align*}
The inner term provides the average value of $x_j$ over the elite solutions that have their binary coordinate $k$ fixed to 1, and ${p}_{\mathrm{corr}}(j)$ averages this probability over $k \in \mathcal{F}_1$. We set the averages to ${1}/{2}$ (neutral) when the denominator is zero. The intuition is that ${p}_{\mathrm{corr}}(j)$ constitutes an (imperfect) adjustment to $p_{\mathrm{base}}(j)$ based on the variables that are currently fixed to 1. We only consider variables that are fixed to $1$ because they are often more informative than those fixed to $0$.
We then decide to fix the variables with the largest value of $|a_\texttt{fix}(j) - 1/2|$.

As a safeguard against potentially time-consuming computation for ${p}_{\mathrm{corr}}(j)$, we fall back to $a_\texttt{fix}(j)=p_{\mathrm{base}}(j)$ if $|\mathcal{F}_1|\cdot|\mathcal{I}_b\setminus(\mathcal{F}_0\cup\mathcal{F}_1)|>5 \cdot 10^6$.

\vspace{12pt}
\paragraph{Time limit.} For \emph{Diving}, \emph{Elite}, and \emph{Deep diving with bound propagation}, we allow for at most 3 seconds per strategy per iteration and fall back to \emph{Reduced costs} when the time budget is exceeded.

\vspace{12pt}
\paragraph{Hyperparameters.} The `Reduced costs' strategy changes the fixing status of $\mu = |\mathcal{J}| / |\mathcal{I}| = 4\%$ of the integer variables. `Diving' fixes $\mu = 5\%$ of the remaining (i.e., unfixed) integer variables. `Consensus restart' defines consensus as shared across 80\% of the elite solutions. `Elite' unfixes and then fixes $\mu = 4\%$ of the binary variables $\mathcal{I}_b$. `Best' fixes 90\% of the coordinates to those of the best incumbent. `Deep diving' does not have a hyperparameter.

For the methods with a hyperparameter denoted by $\mu$ (namely, Reduced Costs, Diving, Elite), we consider an actual rate equal to $\mu$ plus Gaussian noise $\mathcal{N}(0, ({\mu}/{3})^2)$ (with proper clamping), to avoid duplicating the same nodes.

\FloatBarrier

\section{Supplementary Material on the Numerical Experiments}
We now provide supplementary information about the numerical experiments from Section \ref{sec:exp}.

\subsection{Diversifying Node-Generation Strategies (Section \ref{sec:exp:r})}
\label{app:strategy.parallelism}
In this section, we report numerical results supplementing Section~\ref{sec:exp:r}. 

For $R=1$, we use the reduced-cost strategy in Section~\ref{sec:exp:r}. To justify this choice, Figure~\ref{app:fig:strategy.singletons} compares PaNGEA $R=1,W=1$ with the five strategies on the subset of the 25 odd-numbered MIPcc26 instances. The reduced-cost strategy achieves the lowest average gap integral (146.2). 
\begin{figure}
\FIGURE
{\centering
\begin{subfigure}[b]{.48\textwidth}
\centering
\includegraphics[width=\linewidth]{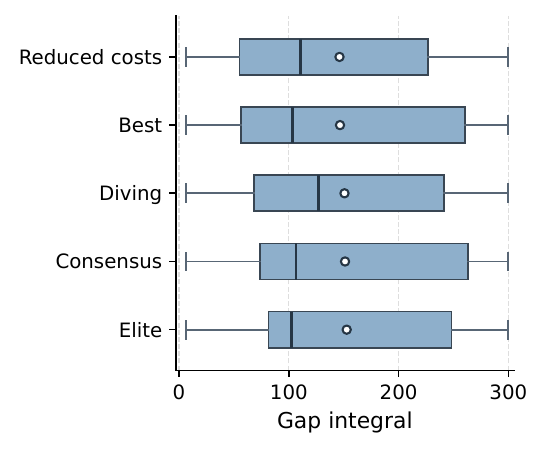}
\caption{Single-strategy distributions.}\label{app:fig:strategy.singletons.a}
\end{subfigure}\hfill
\begin{subfigure}[b]{.48\textwidth}
\centering
\includegraphics[width=\linewidth]{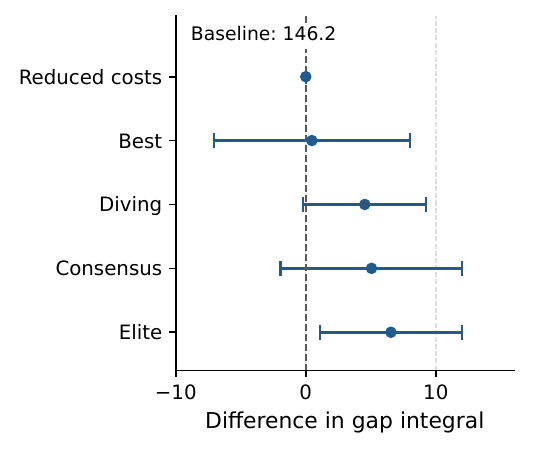}
\caption{Paired differences.}\label{app:fig:strategy.singletons.b}
\end{subfigure}}
{Comparison of single-strategy performance on odd-numbered MIPcc26 instances.
\label{app:fig:strategy.singletons}}
{All configurations use $W=R=1$. In panel (a), boxes show the interquartile range, internal lines show medians, circles show means, and whiskers show the farthest observations within 1.5 interquartile ranges. Panel (b) reports mean within-instance differences relative to baseline (146.2), with $\pm1$ standard error across instances. Negative differences indicate improvement.}
\end{figure}

Figure \ref{fig:W=1.a} reports the performance of PaNGEA with increasing values of $R$, from 1 to 5. We construct these portfolios of strategies greedily: for a fixed value of $R > 1$, we consider the previous $R-1$ strategies, and we add the one strategy that yields the lowest average gap integral on these 25 instances. Table~\ref{app:tab:strategy.greedy} reports the performance of every candidate portfolio examined. The deep diving strategy is aimed at instances on which the root local search fails to find a feasible solution, so we leave it to $R=6$. 

\begin{table}
\TABLE
{All candidates examined along the greedy strategy path.\label{app:tab:strategy.greedy}}
{\small
\setlength{\tabcolsep}{4pt}
\renewcommand{\arraystretch}{1.12}
\begin{tabular}{@{}cp{.62\textwidth}rc@{}}
\toprule
$R$ & Strategies & \shortstack[r]{Average Gap Integral} & Decision \\
\midrule
\textbf{1} & \textbf{Reduced costs} & \textbf{146.18} & \textbf{Selected} \\
1 & Best & 146.65 & -- \\
1 & Diving & 150.72 & -- \\
1 & Consensus & 151.23 & -- \\
1 & Elite & 152.73 & -- \\
\midrule
\textbf{2} & \textbf{Reduced costs + Consensus} & \textbf{147.01} & \textbf{Selected} \\
2 & Reduced costs + Elite & 147.61 & -- \\
2 & Reduced costs + Best & 147.75 & -- \\
2 & Reduced costs + Diving & 149.22 & -- \\
\midrule
\textbf{3} & \textbf{Reduced costs + Consensus + Elite} & \textbf{146.67} & \textbf{Selected} \\
3 & Reduced costs + Consensus + Diving & 149.23 & -- \\
3 & Reduced costs + Consensus + Best & 152.38 & -- \\
\midrule
\textbf{4} & \textbf{Reduced costs + Consensus + Elite + Best} & \textbf{144.19} & \textbf{Selected} \\
4 & Reduced costs + Consensus + Elite + Diving & 159.19 & -- \\
\midrule
\textbf{5} & \textbf{Reduced costs + Consensus + Elite + Best + Diving} & \textbf{143.18} & \textbf{Selected} \\
\midrule
$\textbf{6}^*$ & \textbf{Reduced costs + Consensus + Elite + Best + Diving + Deep diving$^*$} & \textbf{137.56} & \textbf{Selected} \\
\bottomrule
\end{tabular}}
{}
\end{table}

\FloatBarrier
\subsection{Parallel Node Exploration and GPU Utilization (Section \ref{sec:exp:w})}
\label{app:gpu.utilization}

Section \ref{sec:exp:w} investigates the impact of the number of parallel nodes $W$ on PaNGEA's performance, measured in gap integral. We report its impact on GPU utilization in detail in this section.

We report the GPU SM-active rate rather than the \texttt{nvidia-smi} GPU-Util metric. NVIDIA's DCGM documentation defines SM activity as the fraction of time at least one warp is active on a multiprocessor, averaged over all multiprocessors.\endnote{NVIDIA DCGM's SM Activity metric, \url{https://docs.nvidia.com/datacenter/dcgm/latest/user-guide/feature-overview.html}.} In contrast, NVIDIA's \texttt{nvidia-smi} documentation defines GPU-Util as the percent of time over the sampling period during which one or more kernels execute on the GPU.\endnote{NVIDIA System Management Interface documentation, ``Utilization,'' \url{https://docs.nvidia.com/deploy/nvidia-smi/index.html}.} Thus, GPU-Util is a busy-time measure and may be misleading as a proxy for how much of the GPU is occupied, while GPU SM-active rate reflects how broadly work is distributed across the GPU's streaming multiprocessors.

Figures~\ref{app:fig:gpu.util.lp} and~\ref{app:fig:gpu.util.ls} report the performance vs. GPU utilization for PaNGEA with different choices of $W$. The utilization is separated into the relaxation-solving and local-search phases, respectively.
We stratify results across three subsets of instances, based on the number of nonzero entries in the constraint matrix $\bm{A}$.

Across the three subsets of instances, we observe 
that higher {GPU SM-active rate} is generally associated with better final gap; 
that the best fixed value of $W$ varies by subset;
that no single value of $W$ dominates across all instances.

The relaxation-solving phase can terminate early for some nodes, for example, when a relaxation is solved to optimality or proved infeasible. As a result, increasing $W$ does not necessarily increase the effective amount of GPU work throughout the relaxation-solving phase, and the GPU SM-active rate curve need not be monotone in $W$. This is what we observe on the largest instances (in terms of nonzero entries of $\bm{A}$, right panel).
By contrast, local search has no analogous solved-node certificate; once the local-search phase is launched, each node contributes search work until the time (iteration) limit. This matches our observation that local-search GPU SM-active rate increases monotonically with the batch size for all subsets of instances.

The gap results should therefore be read as a performance trade-off rather than as a target to maximize GPU SM-active rate alone. Larger $W$ exposes more parallel node exploration and usually increases GPU activity, especially during local search, but it can also allocate work to less promising nodes. Consequently, the final gap is not monotone in GPU SM-active rate. We observe that our adaptive rule generally achieves higher GPU SM-active rates and better final gaps at the same time.

\begin{figure}
\FIGURE
{\includegraphics[width=\textwidth]{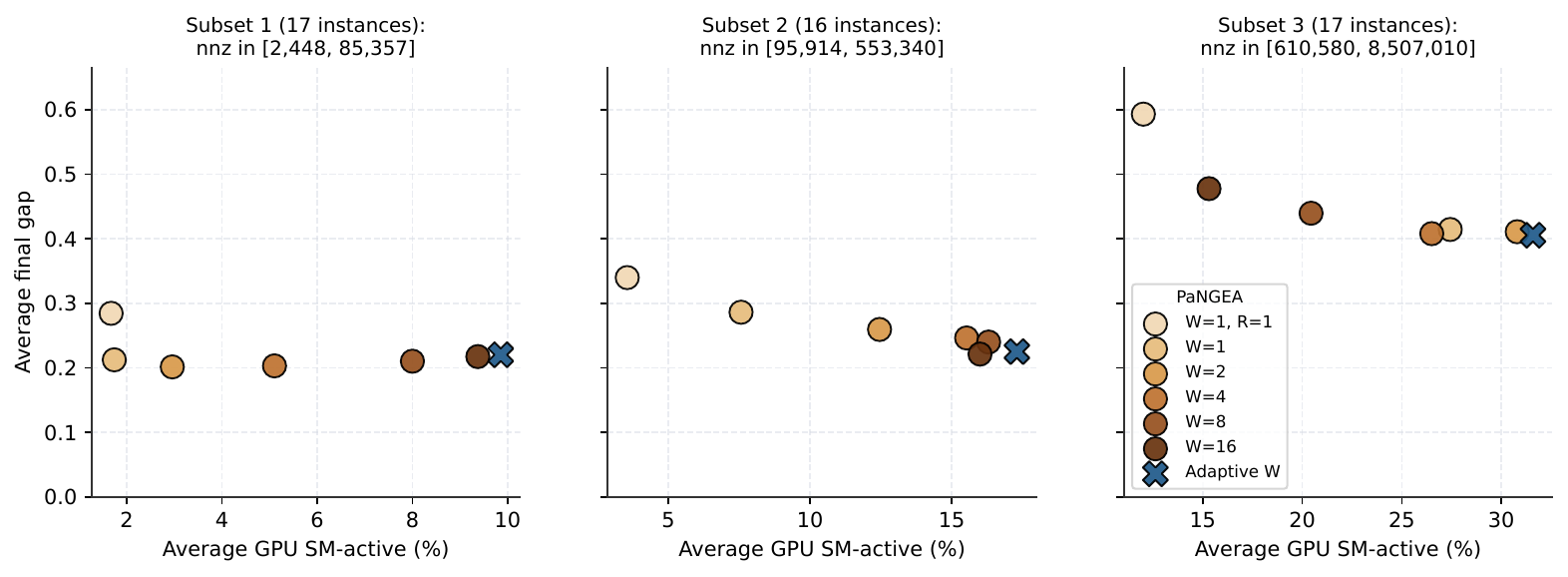}}
{{GPU SM-active rate during solving node relaxations and final gap.} \label{app:fig:gpu.util.lp}}
{The scatter plot shows GPU SM-active rate at the node-relaxation phase vs. final gap for PaNGEA with different values of $W$, averaged within three subsets of MIPcc26 instances grouped by the number of nonzeros in $\bm{A}$.}
\end{figure}

\begin{figure}
\FIGURE
{\includegraphics[width=\textwidth]{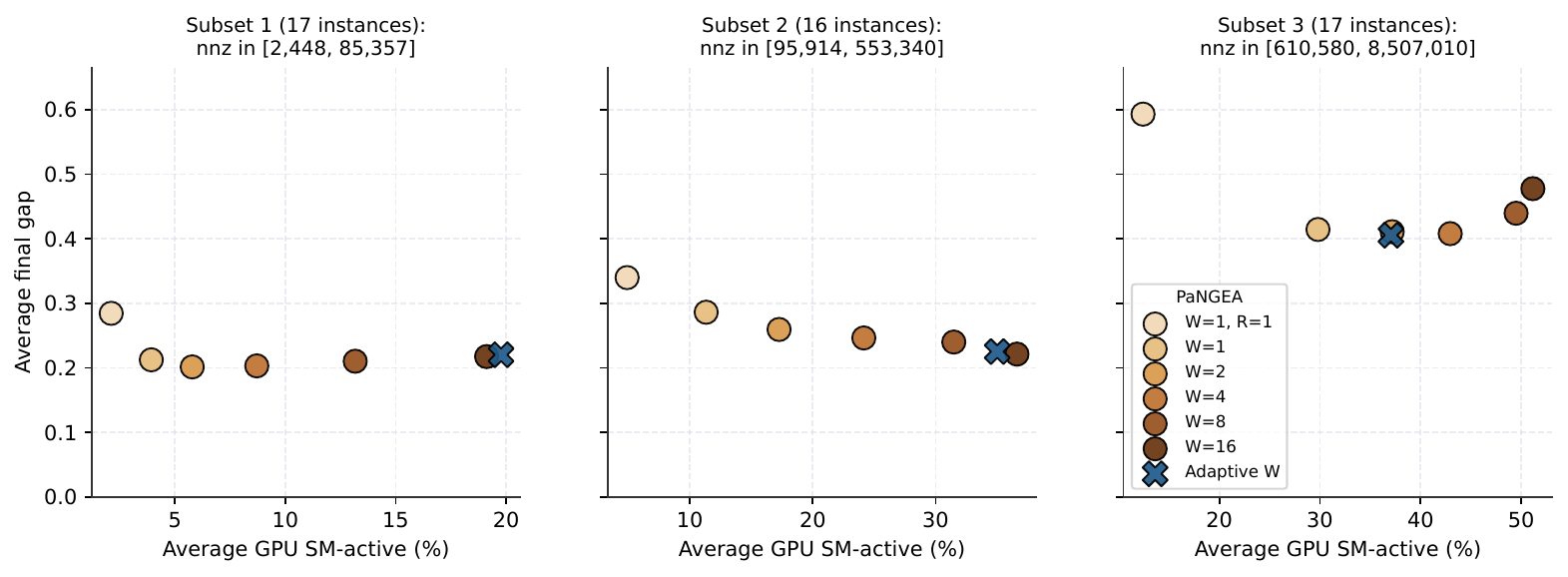}}
{{GPU SM-active rate during local search and final gap.} \label{app:fig:gpu.util.ls}}
{The scatter plot shows GPU SM-active rate at the local-search phase vs.
final gap for PaNGEA with different values of $W$, averaged within three subsets of MIPcc26 instances grouped by the number of nonzeros in $\bm{A}$.}
\end{figure}

\FloatBarrier
\subsection{Benchmarking (Section~\ref{sec:exp:benchmark})}
\label{app:benchmark.supplement}
This section complements the benchmarking diagnostic results from Section~\ref{sec:exp:benchmark}.

Figures~\ref{app:fig:benchmark.boxplot.integral} and \ref{app:fig:benchmark.boxplot.final.gap} report boxplots of the gap integral and final gap distribution, respectively, across the MIPcc26 and MIPLIB instances. 
\begin{figure}
\FIGURE
{\includegraphics[width=.85\textwidth]{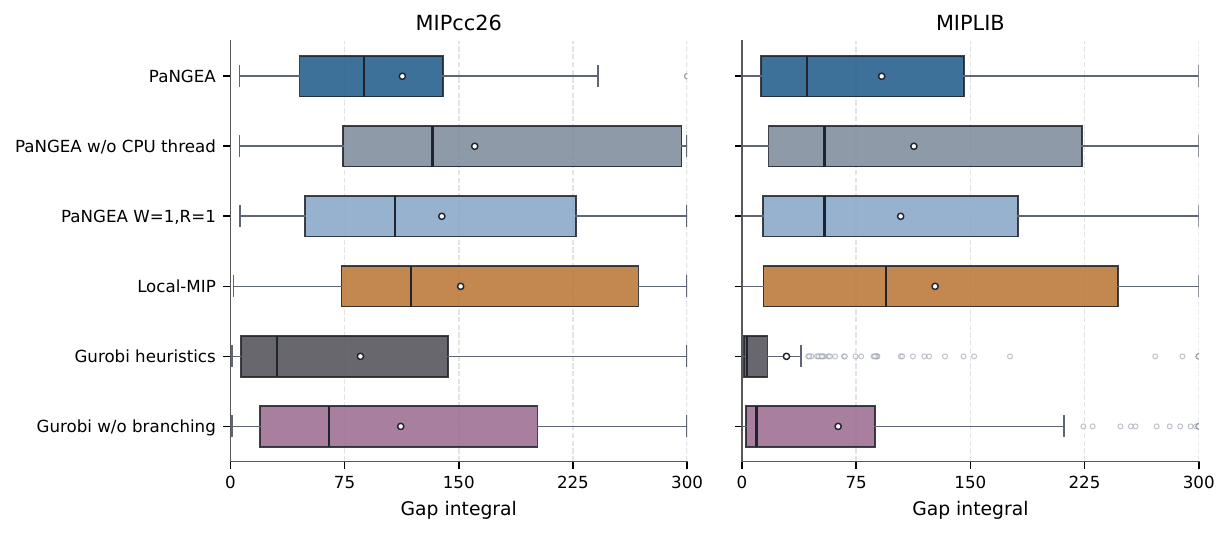}}
{{Gap integral distribution for benchmark methods.} \label{app:fig:benchmark.boxplot.integral}}
{}
\end{figure}
\begin{figure}
\FIGURE
{\includegraphics[width=.85\textwidth]{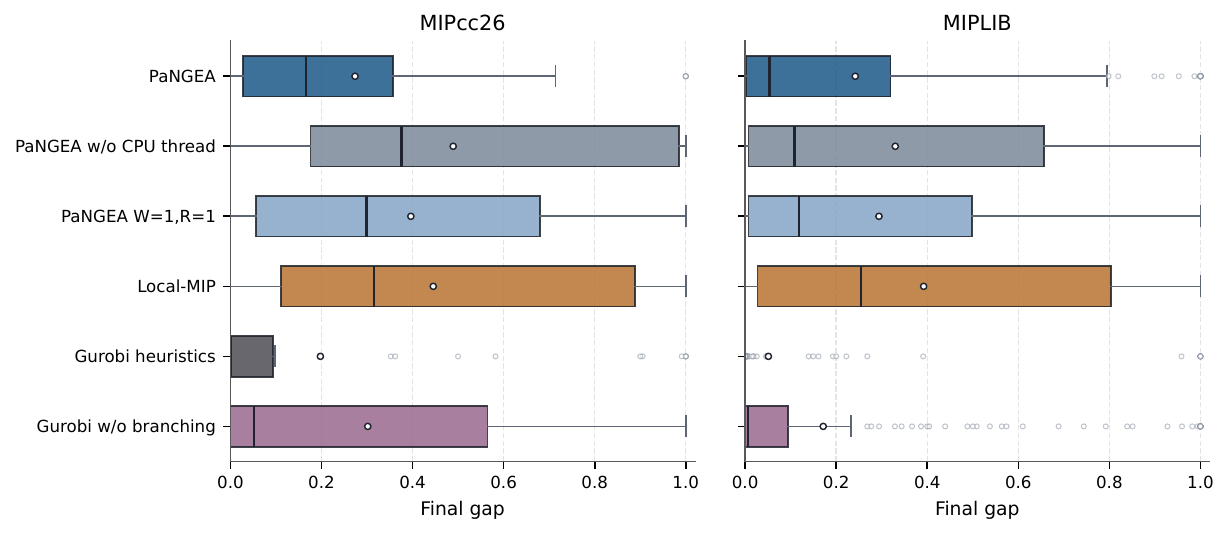}}
{{Final-gap distributions for benchmark methods.} \label{app:fig:benchmark.boxplot.final.gap}}
{}
\end{figure}

\FloatBarrier
Figure~\ref{app:fig:benchmark.instance.size.final.gap} displays the average final gap over all instances (in MIPcc26 and MIPLIB combined) such that $m/n > t$ for increasing values of $t$. 
\begin{figure}
\FIGURE
{\includegraphics[width=.55\textwidth]{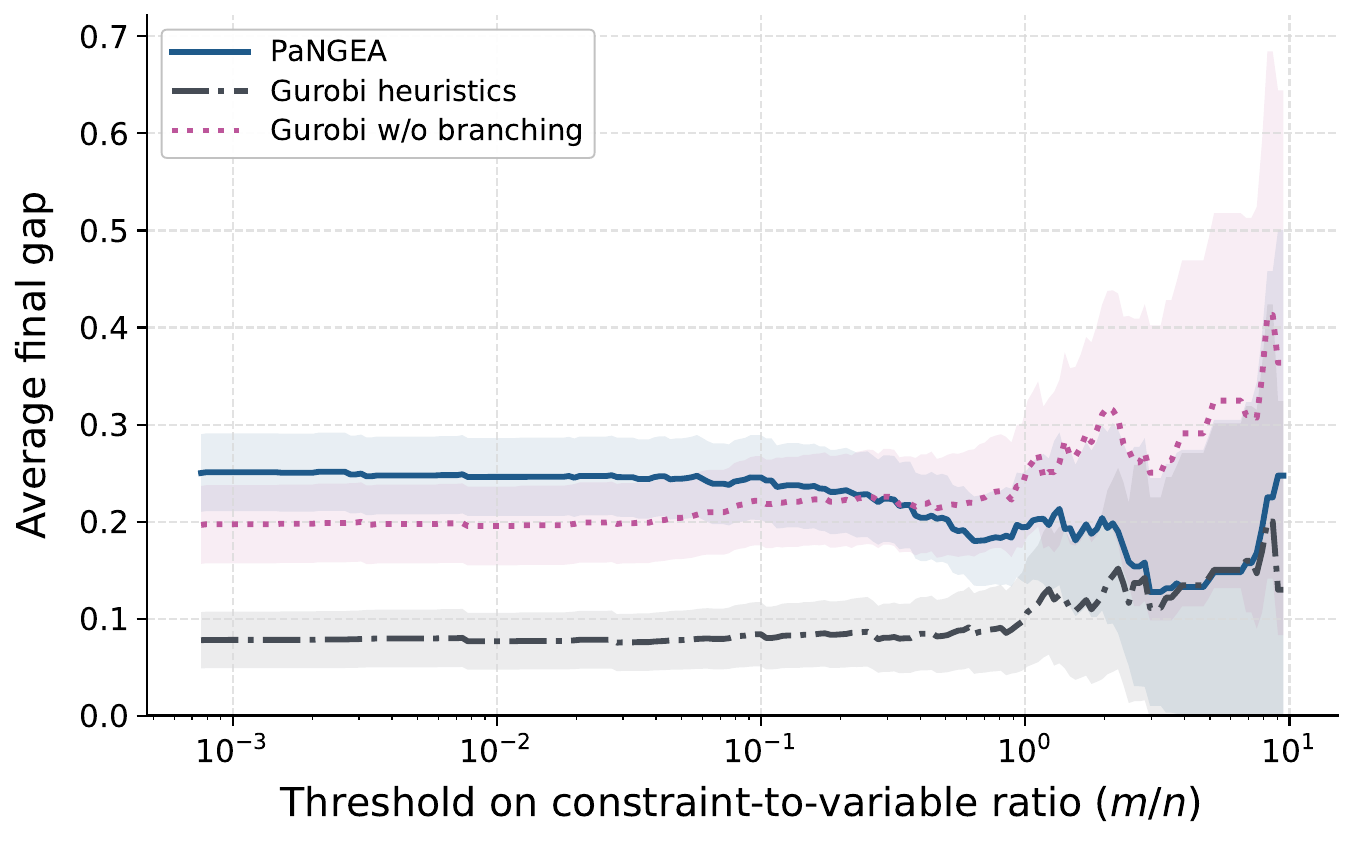}}
{{Average final gap over instances stratified by constraint-to-variable ratio.} \label{app:fig:benchmark.instance.size.final.gap}}
{The x-axis is a threshold $t$ on the constraint-to-variable ratio and is shown on a log scale.  Each line represents the average final gap over all instances with a constraint-to-variable ratio $m/n$ greater than $t$, as $t$ increases. Shaded regions are 95\% confidence intervals.}
\end{figure}

Table~\ref{tab:pangea_win_tie_loss_primal_gap_integral} counts the number of instances where PaNGEA wins/ties/loses against Gurobi's baseline, in terms of gap integral, over all instances and over instances with $m/n>1$. 
The win rate of PaNGEA vs. Gurobi heuristics (resp., Gurobi without branching) increases by around 90\% (resp., 50\%) after narrowing down to instances with $m/n>1$. Table~\ref{tab:pangea_win_tie_loss_final_primal_gap} reports the same analysis, using final gap instead as a metric. 

\begin{table}[t]
\TABLE
{PaNGEA win/tie/loss counts against benchmark comparators in terms of gap integral.\label{tab:pangea_win_tie_loss_primal_gap_integral}}
{\small
\begin{tabular}{lcc}
\toprule
\multirow{2}{*}{PaNGEA vs.} & \multicolumn{2}{c}{Win / Tie / Loss} \\
\cmidrule(lr){2-3}
 & All instances & $m/n>1$ \\
\midrule
Gurobi heuristics & 11 (3.9\%) / 62 / 210 & 8 (7.3\%) / 32 / 69 \\
Gurobi without branching & 73 (25.8\%) / 59 / 151 & 42 (38.5\%) / 26 / 41 \\
\bottomrule
\end{tabular}
}
{}
\end{table}

\begin{table}
\TABLE
{PaNGEA win/tie/loss counts against benchmark comparators in terms of final gap.\label{tab:pangea_win_tie_loss_final_primal_gap}}
{\small
\begin{tabular}{lcc}
\toprule
\multirow{2}{*}{PaNGEA vs.} & \multicolumn{2}{c}{Win / Tie / Loss} \\
\cmidrule(lr){2-3}
 & All instances & $m/n>1$ \\
\midrule
Gurobi heuristics & 30 (10.6\%) / 13 / 240 & 20 (18.3\%) / 5 / 84 \\
Gurobi without branching & 68 (24.0\%) / 18 / 197 & 43 (39.4\%) / 6 / 60 \\
\bottomrule
\end{tabular}
}
{}
\end{table}

\FloatBarrier
\subsection{GPU-CPU Local Search (Section \ref{sec:exp:cpugpu})}
In Figure~\ref{fig:benchmark.cpu.gpu.pattern.examples}, we provide three instance-level examples of distinct contribution patterns. In Figure~\ref{fig:benchmark.cpu.gpu.pattern.examples.a}, the gap is 1 after the root node, and GPU batched local search makes the jump to the first feasible solution, after which CPU-thread local search supplies the remaining improvements. In Figure~\ref{fig:benchmark.cpu.gpu.pattern.examples.b}, CPU Local-MIP accounts for most of the early progress and then stalls, whereas all improvements after 85 seconds come from GPU batched local search. Finally, Figure~\ref{fig:benchmark.cpu.gpu.pattern.examples.c} shows repeated alternation between the two components. These examples illustrate that the role of Local-MIP on the CPU thread is not limited to additional computation and that the interaction of the two components behaves differently depending on the instances.

\begin{figure}
\FIGURE
{\centering
{\begin{subfigure}[b]{.325\hsize}
\centering
\includegraphics[width=\linewidth]{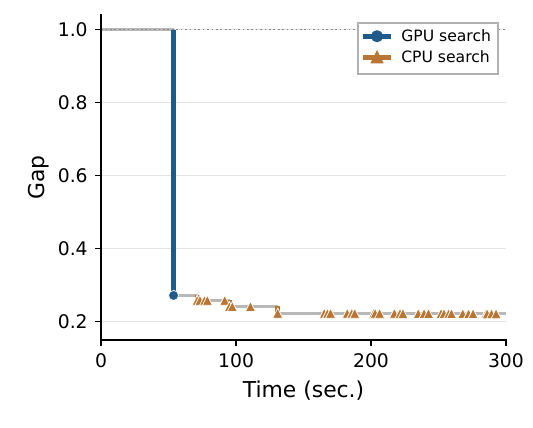}
\caption{MIPcc26: instance-05.\label{fig:benchmark.cpu.gpu.pattern.examples.a}}
\end{subfigure}
\begin{subfigure}[b]{.325\hsize}
\centering
\includegraphics[width=\linewidth]{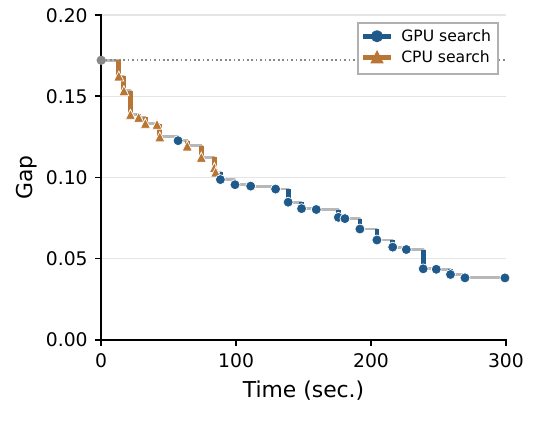}
\caption{MIPcc26: instance-07.\label{fig:benchmark.cpu.gpu.pattern.examples.b}}
\end{subfigure}}
\begin{subfigure}[b]{.325\hsize}
\centering
\includegraphics[width=\linewidth]{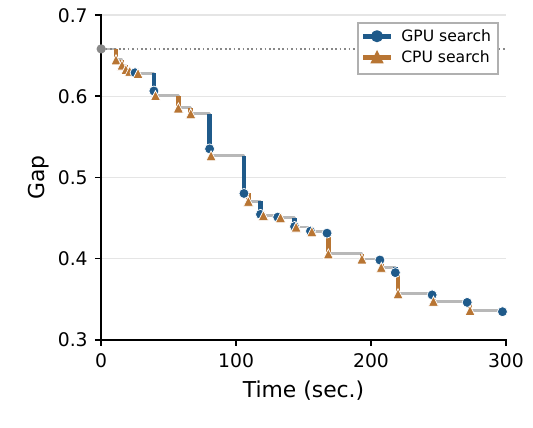}
\caption{MIPcc26: instance-15.\label{fig:benchmark.cpu.gpu.pattern.examples.c}}
\end{subfigure}
}
{Illustrative patterns of CPU Local-MIP vs. GPU batched local search \label{fig:benchmark.cpu.gpu.pattern.examples}}
{Colored gap reductions and markers identify the component producing each incumbent after the root node. }
\end{figure}

\FloatBarrier
\section{Instance-wise Results on the Test Set}
Tables~\ref{app:tab:computational.results.}--\ref{app:tab:computational.results.miplib.5} report the instance-wise feasible objective values of the algorithms benchmarked in Section~\ref{sec:exp:benchmark}.

\begin{table}
\TABLE
{Computational results on the MIPcc26 test set in terms of feasible objective value.\label{app:tab:computational.results.}}
{\setlength{\tabcolsep}{3pt}
\footnotesize
\begin{tabular}{@{}crrrrrrr@{}}
\toprule
\multicolumn{1}{c}{\multirow{2}{*}{Instance}} & \multicolumn{7}{c}{Final Feasible Objective} \\
\cmidrule(lr){2-8}
 & \multicolumn{1}{c}{PaNGEA} & \multicolumn{1}{c}{\shortstack{PaNGEA w/o\\CPU thread}} & \multicolumn{1}{c}{\shortstack{PaNGEA\\W=1,R=1}} & \multicolumn{1}{c}{Local-MIP} & \multicolumn{1}{c}{\shortstack{Gurobi\\heuristics}} & \multicolumn{1}{c}{\shortstack{Gurobi w/o\\branching}} & \multicolumn{1}{c}{\shortstack{Provided\\baseline}} \\
\midrule
01 & $-2.14\mathrm{e}5$ & $-2.14\mathrm{e}5$ & $-2.08\mathrm{e}5$ & $-1.12\mathrm{e}5$ & {$\bm{-2.19\mathrm{e}5}$} & $-2.19\mathrm{e}5$ & $-2.19\mathrm{e}5$ \\
02 & Inf & Inf & Inf & Inf & {$\bm{-3.58\mathrm{e}5}$} & $-3.57\mathrm{e}5$ & $-3.58\mathrm{e}5$ \\
03 & $284.00$ & $837.16$ & $323.60$ & $335.40$ & {$\bm{248.50}$} & $350.10$ & {$\bm{248.50}$} \\
04 & $1.07\mathrm{e}3$ & $1.22\mathrm{e}3$ & Inf & Inf & $882.00$ & {$\bm{868.30}$} & $1.09\mathrm{e}3$ \\
05 & $1.07\mathrm{e}3$ & $1.14\mathrm{e}3$ & Inf & Inf & $922.80$ & {$\bm{833.00}$} & $1.04\mathrm{e}3$ \\
06 & $5.35\mathrm{e}4$ & $5.84\mathrm{e}4$ & $5.40\mathrm{e}4$ & $5.38\mathrm{e}4$ & {$\bm{5.28\mathrm{e}4}$} & $5.63\mathrm{e}4$ & $5.51\mathrm{e}4$ \\
07 & $8.17\mathrm{e}4$ & $8.19\mathrm{e}4$ & $7.95\mathrm{e}4$ & $8.77\mathrm{e}4$ & $7.92\mathrm{e}4$ & {$\bm{7.86\mathrm{e}4}$} & $8.10\mathrm{e}4$ \\
08 & $4.32\mathrm{e}4$ & $4.29\mathrm{e}4$ & $4.50\mathrm{e}4$ & $4.39\mathrm{e}4$ & {$\bm{3.94\mathrm{e}4}$} & $4.07\mathrm{e}4$ & $4.01\mathrm{e}4$ \\
09 & $15.00$ & $16.00$ & $17.00$ & $23.00$ & {$\bm{10.00}$} & $215.00$ & {$\bm{10.00}$} \\
10 & $12.00$ & $11.00$ & $14.00$ & $13.00$ & {$\bm{10.00}$} & $88.00$ & {$\bm{10.00}$} \\
\midrule
11 & $676.00$ & $789.00$ & $687.00$ & $716.00$ & {$\bm{492.00}$} & $603.00$ & $729.00$ \\
12 & $211.00$ & $256.00$ & $207.00$ & $214.00$ & {$\bm{193.00}$} & $198.00$ & {$\bm{193.00}$} \\
13 & Inf & Inf & Inf & Inf & {$\bm{5.00\mathrm{e}3}$} & $5.02\mathrm{e}3$ & $5.05\mathrm{e}3$ \\
14 & $8.63\mathrm{e}5$ & $9.04\mathrm{e}5$ & $1.56\mathrm{e}6$ & $1.56\mathrm{e}6$ & {$\bm{4.92\mathrm{e}5}$} & $4.94\mathrm{e}5$ & $4.95\mathrm{e}5$ \\
15 & $1.02\mathrm{e}6$ & $1.25\mathrm{e}6$ & $1.73\mathrm{e}6$ & $1.69\mathrm{e}6$ & $6.78\mathrm{e}5$ & {$\bm{6.76\mathrm{e}5}$} & $6.78\mathrm{e}5$ \\
16 & $46.00$ & Inf & $74.00$ & {$\bm{45.00}$} & Inf & Inf & Inf \\
17 & Inf & Inf & Inf & Inf & Inf & Inf & Inf \\
18 & $17.00$ & $19.00$ & {$\bm{16.00}$} & $17.00$ & {$\bm{16.00}$} & Inf & {$\bm{16.00}$} \\
19 & $6.00$ & $7.00$ & $6.00$ & $6.00$ & {$\bm{5.00}$} & $6.00$ & {$\bm{5.00}$} \\
20 & $11.00$ & $11.00$ & $11.00$ & $11.00$ & {$\bm{7.00}$} & {$\bm{7.00}$} & {$\bm{7.00}$} \\
\midrule
21 & $12.00$ & $14.00$ & $12.00$ & $14.00$ & {$\bm{10.00}$} & {$\bm{10.00}$} & {$\bm{10.00}$} \\
22 & $228.63$ & $221.63$ & $218.44$ & $250.95$ & {$\bm{171.50}$} & $205.88$ & {$\bm{171.50}$} \\
23 & $263.90$ & $263.90$ & $339.30$ & $403.36$ & {$\bm{100.84}$} & $149.56$ & {$\bm{100.84}$} \\
24 & $436.51$ & $465.02$ & $449.30$ & $462.09$ & {$\bm{431.24}$} & $436.51$ & {$\bm{431.24}$} \\
25 & {$\bm{-32.00}$} & {$\bm{-32.00}$} & {$\bm{-32.00}$} & {$\bm{-32.00}$} & {$\bm{-32.00}$} & $-28.00$ & {$\bm{-32.00}$} \\
26 & {$\bm{-88.00}$} & {$\bm{-88.00}$} & $-87.00$ & $-84.00$ & $-83.00$ & $-84.00$ & $-86.00$ \\
27 & {$\bm{-142.00}$} & $-100.00$ & {$\bm{-142.00}$} & $-100.00$ & $-92.00$ & $-92.00$ & $-92.00$ \\
28 & $2.68\mathrm{e}5$ & $3.88\mathrm{e}5$ & $2.41\mathrm{e}5$ & $2.45\mathrm{e}5$ & {$\bm{2.12\mathrm{e}5}$} & $4.30\mathrm{e}5$ & $3.74\mathrm{e}5$ \\
29 & $8.61\mathrm{e}5$ & Inf & $1.43\mathrm{e}6$ & {$\bm{8.05\mathrm{e}5}$} & Inf & $1.47\mathrm{e}6$ & Inf \\
30 & {$\bm{3.59\mathrm{e}6}$} & Inf & Inf & $4.26\mathrm{e}6$ & Inf & $1.51\mathrm{e}7$ & Inf \\
\midrule
31 & $-2.47\mathrm{e}3$ & $-1.66\mathrm{e}3$ & $-2.46\mathrm{e}3$ & $-2.45\mathrm{e}3$ & $-2.51\mathrm{e}3$ & {$\bm{-2.51\mathrm{e}3}$} & $-2.51\mathrm{e}3$ \\
32 & $-3.41\mathrm{e}3$ & $-1.09\mathrm{e}3$ & $-3.39\mathrm{e}3$ & $-3.36\mathrm{e}3$ & $-3.49\mathrm{e}3$ & {$\bm{-3.49\mathrm{e}3}$} & $-3.49\mathrm{e}3$ \\
33 & {$\bm{-2.04\mathrm{e}3}$} & $-1.62\mathrm{e}3$ & $-872.82$ & $-1.57\mathrm{e}3$ & $-193.65$ & $-193.65$ & $-331.70$ \\
34 & $108.00$ & $120.00$ & $110.00$ & $118.00$ & {$\bm{100.00}$} & $106.00$ & {$\bm{100.00}$} \\
35 & $158.00$ & $176.00$ & $154.00$ & $180.00$ & {$\bm{148.00}$} & $150.00$ & {$\bm{148.00}$} \\
36 & $382.00$ & $530.00$ & $380.00$ & $490.00$ & {$\bm{328.00}$} & {$\bm{328.00}$} & {$\bm{328.00}$} \\
37 & $7.00$ & $26.00$ & $6.00$ & $4.00$ & $4.00$ & Inf & {$\bm{2.00}$} \\
38 & $147.00$ & $3.48\mathrm{e}3$ & $5.30\mathrm{e}4$ & $5.51\mathrm{e}4$ & {$\bm{83.00}$} & $272.00$ & $89.00$ \\
39 & Inf & Inf & Inf & Inf & $478.00$ & {$\bm{464.00}$} & {$\bm{464.00}$} \\
40 & Inf & Inf & Inf & Inf & $570.00$ & {$\bm{524.00}$} & Inf \\
\midrule
41 & Inf & Inf & Inf & Inf & Inf & Inf & Inf \\
42 & $4.12\mathrm{e}7$ & $5.15\mathrm{e}7$ & $2.72\mathrm{e}7$ & $3.78\mathrm{e}8$ & {$\bm{2.49\mathrm{e}7}$} & $2.49\mathrm{e}7$ & $2.49\mathrm{e}7$ \\
43 & $2.85\mathrm{e}7$ & $2.59\mathrm{e}7$ & $3.53\mathrm{e}7$ & $4.12\mathrm{e}8$ & {$\bm{2.43\mathrm{e}7}$} & $2.43\mathrm{e}7$ & $2.43\mathrm{e}7$ \\
44 & $1.36\mathrm{e}8$ & Inf & $1.05\mathrm{e}8$ & $2.12\mathrm{e}9$ & $5.70\mathrm{e}7$ & {$\bm{5.70\mathrm{e}7}$} & $5.70\mathrm{e}7$ \\
45 & {$\bm{4.00}$} & $4.55\mathrm{e}3$ & $23.00$ & $316.00$ & $454.00$ & $8.89\mathrm{e}3$ & {$\bm{4.00}$} \\
46 & {$\bm{1.00}$} & $81.00$ & {$\bm{1.00}$} & {$\bm{1.00}$} & $10.00$ & $180.00$ & $56.00$ \\
47 & $73.00$ & Inf & {$\bm{67.00}$} & $97.00$ & $105.00$ & $128.00$ & $116.00$ \\
48 & $1.90\mathrm{e}8$ & $1.72\mathrm{e}8$ & $2.32\mathrm{e}8$ & $3.00\mathrm{e}8$ & {$\bm{1.52\mathrm{e}8}$} & $1.52\mathrm{e}8$ & $1.52\mathrm{e}8$ \\
49 & {$\bm{1.67\mathrm{e}8}$} & $2.21\mathrm{e}8$ & $2.21\mathrm{e}8$ & $2.78\mathrm{e}8$ & $3.02\mathrm{e}8$ & $3.03\mathrm{e}8$ & $2.69\mathrm{e}8$ \\
50 & $1.92\mathrm{e}8$ & $1.75\mathrm{e}8$ & $2.19\mathrm{e}8$ & $2.82\mathrm{e}8$ & $1.30\mathrm{e}8$ & {$\bm{1.30\mathrm{e}8}$} & $1.30\mathrm{e}8$ \\
\bottomrule
\end{tabular}
}
{}
\end{table}

\begin{table}
\TABLE
{{Computational results on the MIPLIB test set in terms of feasible objective value.}\label{app:tab:computational.results.miplib}}
{\setlength{\tabcolsep}{3pt}
\footnotesize
\begin{tabular}{@{}crrrrrrr@{}}
\toprule
\multicolumn{1}{c}{\multirow{2}{*}{Instance}} & \multicolumn{7}{c}{Final Feasible Objective} \\
\cmidrule(lr){2-8}
 & \multicolumn{1}{c}{PaNGEA} & \multicolumn{1}{c}{\shortstack{PaNGEA w/o\\CPU thread}} & \multicolumn{1}{c}{\shortstack{PaNGEA\\W=1,R=1}} & \multicolumn{1}{c}{Local-MIP} & \multicolumn{1}{c}{\shortstack{Gurobi\\heuristics}} & \multicolumn{1}{c}{\shortstack{Gurobi w/o\\branching}} & \multicolumn{1}{c}{\shortstack{Provided\\baseline}} \\
\midrule
30n20b8 & $402.00$ & $553.00$ & $553.00$ & $553.00$ & $302.00$ & $302.00$ & $302.00$ \\
50v-10 & $3.41\mathrm{e}3$ & $3.50\mathrm{e}3$ & $3.62\mathrm{e}3$ & $4.70\mathrm{e}3$ & $3.31\mathrm{e}3$ & $3.42\mathrm{e}3$ & $3.31\mathrm{e}3$ \\
CMS750\_4 & $252.00$ & $252.00$ & $252.00$ & $252.00$ & $252.00$ & $252.00$ & $252.00$ \\
academictimetablesmall & $681.00$ & $678.00$ & $641.00$ & $1.08\mathrm{e}3$ & $28.00$ & $3.30\mathrm{e}3$ & $0.00$ \\
air05 & $2.65\mathrm{e}4$ & $2.75\mathrm{e}4$ & $2.91\mathrm{e}4$ & $3.44\mathrm{e}4$ & $2.64\mathrm{e}4$ & $2.66\mathrm{e}4$ & $2.64\mathrm{e}4$ \\
app1-1 & $-3.00$ & $-2.00$ & $-2.00$ & Inf & $-3.00$ & $-3.00$ & $-3.00$ \\
app1-2 & Inf & $-20.00$ & Inf & Inf & $-30.00$ & $-30.00$ & $-41.00$ \\
assign1-5-8 & $212.00$ & $212.00$ & $212.00$ & $212.00$ & $212.00$ & $212.00$ & $212.00$ \\
atlanta-ip & $110.01$ & $116.01$ & $126.01$ & $158.01$ & $90.01$ & $93.01$ & $90.01$ \\
b1c1s1 & $3.95\mathrm{e}4$ & $4.70\mathrm{e}4$ & $3.92\mathrm{e}4$ & $4.55\mathrm{e}4$ & $2.45\mathrm{e}4$ & $2.61\mathrm{e}4$ & $2.45\mathrm{e}4$ \\
\midrule
bab2 & Inf & Inf & Inf & Inf & $-3.58\mathrm{e}5$ & $-3.54\mathrm{e}5$ & $-3.58\mathrm{e}5$ \\
bab6 & $-1.93\mathrm{e}5$ & Inf & Inf & Inf & $-2.84\mathrm{e}5$ & $-2.83\mathrm{e}5$ & $-2.84\mathrm{e}5$ \\
beasleyC3 & $781.00$ & $775.00$ & $842.00$ & $850.00$ & $754.00$ & $754.00$ & $754.00$ \\
binkar10\_1 & $6.87\mathrm{e}3$ & $6.80\mathrm{e}3$ & $8.11\mathrm{e}3$ & $9.04\mathrm{e}3$ & $6.74\mathrm{e}3$ & $6.75\mathrm{e}3$ & $6.74\mathrm{e}3$ \\
blp-ar98 & $6.91\mathrm{e}3$ & $6.79\mathrm{e}3$ & $7.11\mathrm{e}3$ & $8.85\mathrm{e}3$ & $6.21\mathrm{e}3$ & $6.21\mathrm{e}3$ & $6.21\mathrm{e}3$ \\
blp-ic98 & $4.98\mathrm{e}3$ & $4.64\mathrm{e}3$ & $5.03\mathrm{e}3$ & $6.05\mathrm{e}3$ & $4.49\mathrm{e}3$ & $4.49\mathrm{e}3$ & $4.49\mathrm{e}3$ \\
bnatt400 & Inf & Inf & Inf & Inf & $1.00$ & Inf & $1.00$ \\
bppc4-08 & $53.00$ & $54.00$ & $53.00$ & $53.00$ & $53.00$ & $56.00$ & $53.00$ \\
brazil3 & $237.00$ & $337.00$ & $227.00$ & $237.00$ & $24.00$ & $331.00$ & $24.00$ \\
buildingenergy & $4.54\mathrm{e}4$ & $5.12\mathrm{e}4$ & $5.19\mathrm{e}4$ & $8.42\mathrm{e}6$ & $3.33\mathrm{e}4$ & $3.33\mathrm{e}4$ & $3.33\mathrm{e}4$ \\
\midrule
cbs-cta & $0.00$ & $43.16$ & $0.00$ & $0.00$ & $0.00$ & $0.00$ & $0.00$ \\
chromaticindex1024-7 & $4.00$ & $4.00$ & $4.00$ & $4.00$ & $4.00$ & $4.00$ & $4.00$ \\
chromaticindex512-7 & $4.00$ & $4.00$ & $4.00$ & $4.00$ & $4.00$ & $4.00$ & $4.00$ \\
cmflsp50-24-8-8 & Inf & Inf & Inf & Inf & $5.69\mathrm{e}7$ & Inf & $5.58\mathrm{e}7$ \\
co-100 & $4.95\mathrm{e}6$ & $6.11\mathrm{e}6$ & $6.08\mathrm{e}6$ & $6.17\mathrm{e}6$ & $2.64\mathrm{e}6$ & $2.64\mathrm{e}6$ & $2.64\mathrm{e}6$ \\
cod105 & $-12.00$ & $-12.00$ & $-12.00$ & $-12.00$ & $-12.00$ & $-12.00$ & $-12.00$ \\
comp07-2idx & $6.00$ & $15.00$ & $6.00$ & $6.00$ & $6.00$ & $10.00$ & $6.00$ \\
comp21-2idx & $113.00$ & $117.00$ & $115.00$ & $113.00$ & $86.00$ & $160.00$ & $74.00$ \\
cost266-UUE & $2.58\mathrm{e}7$ & $2.60\mathrm{e}7$ & $2.92\mathrm{e}7$ & $3.17\mathrm{e}7$ & $2.51\mathrm{e}7$ & $2.65\mathrm{e}7$ & $2.51\mathrm{e}7$ \\
cryptanalysiskb128n5obj16 & Inf & Inf & Inf & Inf & Inf & Inf & $0.00$ \\
\midrule
csched007 & Inf & Inf & Inf & Inf & $351.00$ & $572.00$ & $351.00$ \\
csched008 & $179.00$ & $183.00$ & $179.00$ & $179.00$ & $173.00$ & $186.00$ & $173.00$ \\
cvs16r128-89 & $-95.00$ & $-91.00$ & $-96.00$ & $-94.00$ & $-97.00$ & $-95.00$ & $-97.00$ \\
dano3\_3 & $623.21$ & $897.25$ & $634.54$ & $622.97$ & $576.34$ & $576.34$ & $576.34$ \\
dano3\_5 & $660.41$ & Inf & $647.15$ & $655.56$ & $576.92$ & $576.92$ & $576.92$ \\
decomp2 & $-160.00$ & $-160.00$ & $-160.00$ & $-160.00$ & $-160.00$ & $-160.00$ & $-160.00$ \\
drayage-100-23 & $1.03\mathrm{e}5$ & $1.03\mathrm{e}5$ & $1.03\mathrm{e}5$ & $1.83\mathrm{e}5$ & $1.03\mathrm{e}5$ & $1.03\mathrm{e}5$ & $1.03\mathrm{e}5$ \\
drayage-25-23 & $1.01\mathrm{e}5$ & $1.01\mathrm{e}5$ & $1.07\mathrm{e}5$ & $1.71\mathrm{e}5$ & $1.01\mathrm{e}5$ & $1.01\mathrm{e}5$ & $1.01\mathrm{e}5$ \\
dws008-01 & Inf & Inf & Inf & Inf & $3.74\mathrm{e}4$ & $7.62\mathrm{e}4$ & $3.74\mathrm{e}4$ \\
eil33-2 & $934.01$ & $988.14$ & $974.66$ & $989.39$ & $934.01$ & $1.07\mathrm{e}3$ & $934.01$ \\
\midrule
eilA101-2 & $1.43\mathrm{e}3$ & $2.41\mathrm{e}3$ & $1.46\mathrm{e}3$ & $1.38\mathrm{e}3$ & $923.01$ & $923.01$ & $880.92$ \\
enlight\_hard & $37.00$ & $37.00$ & $37.00$ & $37.00$ & $37.00$ & $37.00$ & $37.00$ \\
ex10 & $100.00$ & $100.00$ & $100.00$ & $100.00$ & $100.00$ & $100.00$ & $100.00$ \\
ex9 & $81.00$ & $81.00$ & $81.00$ & $81.00$ & $81.00$ & $81.00$ & $81.00$ \\
exp-1-500-5-5 & $7.74\mathrm{e}4$ & $7.40\mathrm{e}4$ & $1.17\mathrm{e}5$ & $1.41\mathrm{e}5$ & $6.59\mathrm{e}4$ & $6.72\mathrm{e}4$ & $6.59\mathrm{e}4$ \\
fast0507 & $175.00$ & $175.00$ & $183.00$ & $191.00$ & $174.00$ & $175.00$ & $174.00$ \\
fastxgemm-n2r6s0t2 & $230.00$ & $230.00$ & $230.00$ & $230.00$ & $230.00$ & $527.00$ & $230.00$ \\
fhnw-binpack4-48 & $0.00$ & $0.00$ & $0.00$ & $0.00$ & $0.00$ & Inf & $0.00$ \\
fiball & $138.00$ & $138.00$ & $138.00$ & $138.00$ & $138.00$ & $138.00$ & $138.00$ \\
gen-ip002 & $-4.75\mathrm{e}3$ & $-4.75\mathrm{e}3$ & $-4.72\mathrm{e}3$ & $-4.48\mathrm{e}3$ & $-4.78\mathrm{e}3$ & $-4.75\mathrm{e}3$ & $-4.78\mathrm{e}3$ \\
\bottomrule
\end{tabular}
}
{}
\end{table}

\begin{table}
\TABLE
{{Computational results on the MIPLIB test set in terms of feasible objective value (continued).}\label{app:tab:computational.results.miplib.2}}
{\setlength{\tabcolsep}{3pt}
\footnotesize
\begin{tabular}{@{}crrrrrrr@{}}
\toprule
\multicolumn{1}{c}{\multirow{2}{*}{Instance}} & \multicolumn{7}{c}{Final Feasible Objective} \\
\cmidrule(lr){2-8}
 & \multicolumn{1}{c}{PaNGEA} & \multicolumn{1}{c}{\shortstack{PaNGEA w/o\\CPU thread}} & \multicolumn{1}{c}{\shortstack{PaNGEA\\W=1,R=1}} & \multicolumn{1}{c}{Local-MIP} & \multicolumn{1}{c}{\shortstack{Gurobi\\heuristics}} & \multicolumn{1}{c}{\shortstack{Gurobi w/o\\branching}} & \multicolumn{1}{c}{\shortstack{Provided\\baseline}} \\
\midrule
gen-ip054 & $6.90\mathrm{e}3$ & $6.87\mathrm{e}3$ & $6.89\mathrm{e}3$ & $6.92\mathrm{e}3$ & $6.84\mathrm{e}3$ & $6.90\mathrm{e}3$ & $6.84\mathrm{e}3$ \\
germanrr & $5.23\mathrm{e}7$ & $5.36\mathrm{e}7$ & $5.24\mathrm{e}7$ & $5.32\mathrm{e}7$ & $4.72\mathrm{e}7$ & $4.71\mathrm{e}7$ & $4.71\mathrm{e}7$ \\
gfd-schedulen180f7d50m30k18 & Inf & Inf & Inf & Inf & Inf & Inf & $1.00$ \\
glass-sc & $23.00$ & $23.00$ & $23.00$ & $23.00$ & $23.00$ & $23.00$ & $23.00$ \\
glass4 & $1.45\mathrm{e}9$ & $1.58\mathrm{e}9$ & $1.60\mathrm{e}9$ & $1.70\mathrm{e}9$ & $1.20\mathrm{e}9$ & $1.70\mathrm{e}9$ & $1.20\mathrm{e}9$ \\
gmu-35-40 & $-2.40\mathrm{e}6$ & $-2.40\mathrm{e}6$ & $-2.39\mathrm{e}6$ & $-2.34\mathrm{e}6$ & $-2.41\mathrm{e}6$ & $-2.41\mathrm{e}6$ & $-2.41\mathrm{e}6$ \\
gmu-35-50 & $-2.60\mathrm{e}6$ & $-2.60\mathrm{e}6$ & $-2.56\mathrm{e}6$ & $-2.56\mathrm{e}6$ & $-2.61\mathrm{e}6$ & $-2.61\mathrm{e}6$ & $-2.61\mathrm{e}6$ \\
graph20-20-1rand & $-9.00$ & $-9.00$ & $-9.00$ & $-9.00$ & $-9.00$ & $-9.00$ & $-9.00$ \\
graphdraw-domain & $1.98\mathrm{e}4$ & $1.97\mathrm{e}4$ & $1.99\mathrm{e}4$ & $2.10\mathrm{e}4$ & $1.97\mathrm{e}4$ & $2.55\mathrm{e}4$ & $1.97\mathrm{e}4$ \\
h80x6320d & $8.24\mathrm{e}3$ & $8.24\mathrm{e}3$ & $8.24\mathrm{e}3$ & $1.61\mathrm{e}4$ & $6.38\mathrm{e}3$ & $6.38\mathrm{e}3$ & $6.38\mathrm{e}3$ \\
\midrule
highschool1-aigio & $1.05\mathrm{e}3$ & $1.38\mathrm{e}3$ & $1.33\mathrm{e}3$ & $1.81\mathrm{e}3$ & $28.00$ & Inf & $0.00$ \\
hypothyroid-k1 & $-2.85\mathrm{e}3$ & $-2.85\mathrm{e}3$ & $-2.85\mathrm{e}3$ & $-2.85\mathrm{e}3$ & $-2.85\mathrm{e}3$ & $-2.85\mathrm{e}3$ & $-2.85\mathrm{e}3$ \\
ic97\_potential & $3.98\mathrm{e}3$ & $4.00\mathrm{e}3$ & $4.04\mathrm{e}3$ & $4.05\mathrm{e}3$ & $3.94\mathrm{e}3$ & $4.05\mathrm{e}3$ & $3.94\mathrm{e}3$ \\
icir97\_tension & $6.64\mathrm{e}3$ & Inf & Inf & $6.99\mathrm{e}3$ & $6.38\mathrm{e}3$ & Inf & $6.38\mathrm{e}3$ \\
irish-electricity & $6.04\mathrm{e}6$ & Inf & $5.70\mathrm{e}6$ & $6.18\mathrm{e}6$ & $3.72\mathrm{e}6$ & $3.72\mathrm{e}6$ & $3.72\mathrm{e}6$ \\
irp & $1.22\mathrm{e}4$ & $1.22\mathrm{e}4$ & $1.22\mathrm{e}4$ & $1.22\mathrm{e}4$ & $1.22\mathrm{e}4$ & $1.22\mathrm{e}4$ & $1.22\mathrm{e}4$ \\
istanbul-no-cutoff & $257.94$ & Inf & Inf & Inf & $204.08$ & $204.08$ & $204.08$ \\
k1mushroom & $-3.29\mathrm{e}3$ & $-756.00$ & $-3.29\mathrm{e}3$ & $-3.29\mathrm{e}3$ & $-3.29\mathrm{e}3$ & $-3.29\mathrm{e}3$ & $-3.29\mathrm{e}3$ \\
lectsched-5-obj & $24.00$ & $26.00$ & $24.00$ & $24.00$ & $24.00$ & $27.00$ & $24.00$ \\
leo1 & $4.52\mathrm{e}8$ & $4.15\mathrm{e}8$ & $4.69\mathrm{e}8$ & $5.01\mathrm{e}8$ & $4.05\mathrm{e}8$ & $4.08\mathrm{e}8$ & $4.04\mathrm{e}8$ \\
\midrule
leo2 & $4.49\mathrm{e}8$ & $4.39\mathrm{e}8$ & $5.42\mathrm{e}8$ & $5.50\mathrm{e}8$ & $4.06\mathrm{e}8$ & $4.06\mathrm{e}8$ & $4.04\mathrm{e}8$ \\
lotsize & $2.66\mathrm{e}6$ & $2.54\mathrm{e}6$ & $3.32\mathrm{e}6$ & $3.77\mathrm{e}6$ & $1.48\mathrm{e}6$ & $1.53\mathrm{e}6$ & $1.48\mathrm{e}6$ \\
mad & $0.31$ & $0.34$ & $0.33$ & $0.52$ & $0.04$ & $0.66$ & $0.03$ \\
map10 & $-475.00$ & $-430.00$ & $-481.00$ & $-414.00$ & $-495.00$ & $-478.00$ & $-495.00$ \\
map16715-04 & $-94.00$ & $-91.00$ & $-70.00$ & $-55.10$ & $-111.00$ & $-108.00$ & $-111.00$ \\
markshare2 & $75.00$ & $105.00$ & $114.00$ & $107.00$ & $24.00$ & $162.00$ & $1.00$ \\
markshare\_4\_0 & $3.00$ & $6.00$ & $3.00$ & $3.00$ & $1.00$ & $56.00$ & $1.00$ \\
mas74 & $1.23\mathrm{e}4$ & $1.30\mathrm{e}4$ & $1.33\mathrm{e}4$ & $1.98\mathrm{e}4$ & $1.19\mathrm{e}4$ & $1.25\mathrm{e}4$ & $1.18\mathrm{e}4$ \\
mas76 & $4.10\mathrm{e}4$ & $4.09\mathrm{e}4$ & $4.16\mathrm{e}4$ & $4.67\mathrm{e}4$ & $4.00\mathrm{e}4$ & $4.00\mathrm{e}4$ & $4.00\mathrm{e}4$ \\
mc11 & $1.19\mathrm{e}4$ & $1.20\mathrm{e}4$ & $1.35\mathrm{e}4$ & $2.94\mathrm{e}4$ & $1.17\mathrm{e}4$ & $1.17\mathrm{e}4$ & $1.17\mathrm{e}4$ \\
\midrule
mcsched & $2.14\mathrm{e}5$ & $2.15\mathrm{e}5$ & $2.14\mathrm{e}5$ & $2.16\mathrm{e}5$ & $2.12\mathrm{e}5$ & $2.14\mathrm{e}5$ & $2.12\mathrm{e}5$ \\
mik-250-20-75-4 & $-5.19\mathrm{e}4$ & $-5.19\mathrm{e}4$ & $-4.47\mathrm{e}4$ & $-4.24\mathrm{e}4$ & $-5.23\mathrm{e}4$ & $-5.23\mathrm{e}4$ & $-5.23\mathrm{e}4$ \\
milo-v12-6-r2-40-1 & $1.24\mathrm{e}6$ & $1.24\mathrm{e}6$ & Inf & Inf & $3.26\mathrm{e}5$ & $7.66\mathrm{e}5$ & $3.26\mathrm{e}5$ \\
momentum1 & Inf & Inf & Inf & Inf & $1.09\mathrm{e}5$ & $1.09\mathrm{e}5$ & $1.09\mathrm{e}5$ \\
mushroom-best & $0.06$ & $0.06$ & $0.06$ & $0.06$ & $0.06$ & $0.06$ & $0.06$ \\
mzzv11 & $-1.90\mathrm{e}4$ & $-1.98\mathrm{e}4$ & $-2.00\mathrm{e}4$ & $-1.85\mathrm{e}4$ & $-2.17\mathrm{e}4$ & $-2.17\mathrm{e}4$ & $-2.17\mathrm{e}4$ \\
mzzv42z & $-2.01\mathrm{e}4$ & $-1.95\mathrm{e}4$ & $-2.01\mathrm{e}4$ & $-1.97\mathrm{e}4$ & $-2.05\mathrm{e}4$ & $-2.05\mathrm{e}4$ & $-2.05\mathrm{e}4$ \\
n2seq36q & $5.88\mathrm{e}4$ & $8.48\mathrm{e}4$ & $5.38\mathrm{e}4$ & $9.50\mathrm{e}4$ & $5.22\mathrm{e}4$ & $5.22\mathrm{e}4$ & $5.22\mathrm{e}4$ \\
n3div36 & $1.31\mathrm{e}5$ & $1.31\mathrm{e}5$ & $1.49\mathrm{e}5$ & $1.47\mathrm{e}5$ & $1.31\mathrm{e}5$ & $1.31\mathrm{e}5$ & $1.31\mathrm{e}5$ \\
n5-3 & $9.52\mathrm{e}3$ & $1.03\mathrm{e}4$ & $1.09\mathrm{e}4$ & $1.32\mathrm{e}4$ & $8.10\mathrm{e}3$ & $8.14\mathrm{e}3$ & $8.10\mathrm{e}3$ \\
\midrule
neos-1122047 & $161.00$ & $161.00$ & $161.00$ & $161.00$ & $161.00$ & $161.00$ & $161.00$ \\
neos-1171448 & $-309.00$ & $-309.00$ & $-309.00$ & $-299.00$ & $-309.00$ & $-309.00$ & $-309.00$ \\
neos-1171737 & $-193.00$ & $-193.00$ & $-192.00$ & $-190.21$ & $-195.00$ & $-195.00$ & $-195.00$ \\
neos-1354092 & Inf & Inf & Inf & Inf & $57.00$ & $46.00$ & $46.00$ \\
neos-1445765 & $-1.74\mathrm{e}4$ & $-1.74\mathrm{e}4$ & $-1.54\mathrm{e}4$ & $-1.33\mathrm{e}4$ & $-1.78\mathrm{e}4$ & $-1.78\mathrm{e}4$ & $-1.78\mathrm{e}4$ \\
neos-1456979 & $263.00$ & $315.00$ & $351.00$ & $357.00$ & $176.00$ & $451.00$ & $176.00$ \\
neos-1582420 & $102.00$ & $103.00$ & $129.00$ & $153.00$ & $91.00$ & $95.00$ & $91.00$ \\
neos-2657525-crna & $5.68$ & $7.23$ & $1.81$ & $5.77$ & $1.81$ & $12.15$ & $1.81$ \\
neos-2746589-doon & Inf & Inf & Inf & Inf & $2.01\mathrm{e}3$ & $3.58\mathrm{e}3$ & $2.01\mathrm{e}3$ \\
neos-2978193-inde & $-0.67$ & $1.91$ & $4.54$ & $13.54$ & $-2.39$ & $-2.39$ & $-2.39$ \\
\bottomrule
\end{tabular}
}
{}
\end{table}

\begin{table}
\TABLE
{{Computational results on the MIPLIB test set in terms of feasible objective value (continued).}\label{app:tab:computational.results.miplib.3}}
{\setlength{\tabcolsep}{3pt}
\footnotesize
\begin{tabular}{@{}crrrrrrr@{}}
\toprule
\multicolumn{1}{c}{\multirow{2}{*}{Instance}} & \multicolumn{7}{c}{Final Feasible Objective} \\
\cmidrule(lr){2-8}
 & \multicolumn{1}{c}{PaNGEA} & \multicolumn{1}{c}{\shortstack{PaNGEA w/o\\CPU thread}} & \multicolumn{1}{c}{\shortstack{PaNGEA\\W=1,R=1}} & \multicolumn{1}{c}{Local-MIP} & \multicolumn{1}{c}{\shortstack{Gurobi\\heuristics}} & \multicolumn{1}{c}{\shortstack{Gurobi w/o\\branching}} & \multicolumn{1}{c}{\shortstack{Provided\\baseline}} \\
\midrule
neos-2987310-joes & $-5.75\mathrm{e}8$ & $-3.64\mathrm{e}8$ & $-5.96\mathrm{e}8$ & $-5.39\mathrm{e}8$ & $-6.08\mathrm{e}8$ & $-6.08\mathrm{e}8$ & $-6.08\mathrm{e}8$ \\
neos-3004026-krka & $0.00$ & $0.00$ & $0.00$ & $0.00$ & $0.00$ & $0.00$ & $0.00$ \\
neos-3024952-loue & $6.06\mathrm{e}4$ & $5.62\mathrm{e}4$ & $1.12\mathrm{e}5$ & $1.50\mathrm{e}5$ & $2.68\mathrm{e}4$ & Inf & $2.68\mathrm{e}4$ \\
neos-3046615-murg & $1.63\mathrm{e}3$ & $1.62\mathrm{e}3$ & $1.69\mathrm{e}3$ & $1.75\mathrm{e}3$ & $1.60\mathrm{e}3$ & $1.64\mathrm{e}3$ & $1.60\mathrm{e}3$ \\
neos-3083819-nubu & $6.32\mathrm{e}6$ & $6.32\mathrm{e}6$ & $6.37\mathrm{e}6$ & $6.49\mathrm{e}6$ & $6.31\mathrm{e}6$ & $6.31\mathrm{e}6$ & $6.31\mathrm{e}6$ \\
neos-3216931-puriri & $1.51\mathrm{e}5$ & Inf & $1.31\mathrm{e}5$ & $1.56\mathrm{e}5$ & $7.13\mathrm{e}4$ & Inf & $7.13\mathrm{e}4$ \\
neos-3381206-awhea & $473.00$ & Inf & $471.00$ & Inf & $453.00$ & $453.00$ & $453.00$ \\
neos-3402294-bobin & $0.07$ & $0.07$ & $0.07$ & $0.07$ & $0.07$ & $0.07$ & $0.07$ \\
neos-3555904-turama & $-34.70$ & $-34.70$ & $-34.70$ & $-34.70$ & $-34.70$ & $-34.70$ & $-34.70$ \\
neos-3627168-kasai & $9.89\mathrm{e}5$ & $9.89\mathrm{e}5$ & $9.95\mathrm{e}5$ & $1.00\mathrm{e}6$ & $9.89\mathrm{e}5$ & $9.89\mathrm{e}5$ & $9.89\mathrm{e}5$ \\
\midrule
neos-3656078-kumeu & $-1.30\mathrm{e}4$ & $-1.24\mathrm{e}4$ & $-1.29\mathrm{e}4$ & $-1.29\mathrm{e}4$ & $-1.32\mathrm{e}4$ & Inf & $-1.32\mathrm{e}4$ \\
neos-3754480-nidda & $1.30\mathrm{e}4$ & $1.30\mathrm{e}4$ & $1.51\mathrm{e}4$ & $1.66\mathrm{e}4$ & $1.29\mathrm{e}4$ & $1.45\mathrm{e}4$ & $1.29\mathrm{e}4$ \\
neos-4300652-rahue & $3.45$ & $6.23$ & $3.55$ & $4.74$ & $2.14$ & $2.14$ & $2.14$ \\
neos-4338804-snowy & $1.48\mathrm{e}3$ & $1.48\mathrm{e}3$ & $1.48\mathrm{e}3$ & $1.49\mathrm{e}3$ & $1.48\mathrm{e}3$ & $1.59\mathrm{e}3$ & $1.47\mathrm{e}3$ \\
neos-4387871-tavua & $34.38$ & $36.38$ & $47.87$ & $86.67$ & $33.38$ & $34.56$ & $33.38$ \\
neos-4413714-turia & $56.04$ & Inf & Inf & Inf & $45.37$ & $45.37$ & $45.37$ \\
neos-4532248-waihi & $61.60$ & $65.50$ & $74.70$ & Inf & $61.60$ & $69.80$ & $61.60$ \\
neos-4647030-tutaki & $3.64\mathrm{e}4$ & $4.33\mathrm{e}4$ & $4.35\mathrm{e}4$ & $6.39\mathrm{e}4$ & $2.73\mathrm{e}4$ & $2.73\mathrm{e}4$ & $2.73\mathrm{e}4$ \\
neos-4722843-widden & $2.54\mathrm{e}4$ & $2.54\mathrm{e}4$ & Inf & Inf & $2.50\mathrm{e}4$ & $2.56\mathrm{e}4$ & $2.50\mathrm{e}4$ \\
neos-4738912-atrato & $2.99\mathrm{e}8$ & $2.99\mathrm{e}8$ & $3.22\mathrm{e}8$ & $4.17\mathrm{e}8$ & $2.84\mathrm{e}8$ & $2.85\mathrm{e}8$ & $2.84\mathrm{e}8$ \\
\midrule
neos-4763324-toguru & $1.71\mathrm{e}3$ & $1.80\mathrm{e}3$ & $1.83\mathrm{e}3$ & $2.58\mathrm{e}3$ & $1.61\mathrm{e}3$ & $1.61\mathrm{e}3$ & $1.61\mathrm{e}3$ \\
neos-4954672-berkel & $2.76\mathrm{e}6$ & $2.98\mathrm{e}6$ & $3.55\mathrm{e}6$ & $4.01\mathrm{e}6$ & $2.61\mathrm{e}6$ & $2.72\mathrm{e}6$ & $2.61\mathrm{e}6$ \\
neos-5049753-cuanza & $672.00$ & Inf & $1.71\mathrm{e}3$ & $2.08\mathrm{e}3$ & $562.00$ & $573.00$ & $562.00$ \\
neos-5052403-cygnet & $188.00$ & $191.00$ & $183.00$ & $191.00$ & $182.00$ & $182.00$ & $182.00$ \\
neos-5093327-huahum & Inf & Inf & Inf & Inf & $6.26\mathrm{e}3$ & $3.89\mathrm{e}4$ & $6.26\mathrm{e}3$ \\
neos-5104907-jarama & Inf & Inf & Inf & $2.68\mathrm{e}3$ & Inf & $946.00$ & $935.00$ \\
neos-5107597-kakapo & $4.45\mathrm{e}3$ & $1.87\mathrm{e}4$ & $5.83\mathrm{e}3$ & $2.19\mathrm{e}4$ & $3.64\mathrm{e}3$ & $1.42\mathrm{e}4$ & $3.64\mathrm{e}3$ \\
neos-5114902-kasavu & Inf & Inf & $2.58\mathrm{e}3$ & $1.97\mathrm{e}3$ & $657.00$ & $660.00$ & $655.00$ \\
neos-5188808-nattai & $0.16$ & $2.34$ & $0.12$ & $0.14$ & $0.11$ & $0.22$ & $0.11$ \\
neos-5195221-niemur & $0.01$ & $0.01$ & $0.01$ & $0.01$ & $0.00$ & $0.01$ & $0.00$ \\
\midrule
neos-631710 & $206.00$ & $212.00$ & $205.00$ & $206.00$ & $203.00$ & $203.00$ & $203.00$ \\
neos-662469 & $2.35\mathrm{e}5$ & $2.35\mathrm{e}5$ & $4.85\mathrm{e}5$ & $1.35\mathrm{e}6$ & $1.84\mathrm{e}5$ & $1.84\mathrm{e}5$ & $1.84\mathrm{e}5$ \\
neos-787933 & $30.00$ & $30.00$ & $30.00$ & $30.00$ & $30.00$ & $30.00$ & $30.00$ \\
neos-827175 & $112.00$ & $112.00$ & $112.00$ & $112.00$ & $112.00$ & $112.00$ & $112.00$ \\
neos-848589 & $1.31\mathrm{e}6$ & $2.51\mathrm{e}7$ & $1.64\mathrm{e}7$ & $3.11\mathrm{e}7$ & $2.35\mathrm{e}3$ & $2.53\mathrm{e}3$ & $2.35\mathrm{e}3$ \\
neos-860300 & $3.48\mathrm{e}3$ & $3.46\mathrm{e}3$ & $3.69\mathrm{e}3$ & Inf & $3.20\mathrm{e}3$ & $3.54\mathrm{e}3$ & $3.20\mathrm{e}3$ \\
neos-873061 & $425.78$ & $5.43\mathrm{e}3$ & $138.67$ & $1.61\mathrm{e}4$ & $113.66$ & $113.66$ & $113.66$ \\
neos-911970 & $56.05$ & $55.85$ & $59.31$ & $64.21$ & $54.76$ & $55.26$ & $54.76$ \\
neos-933966 & $342.00$ & $1.34\mathrm{e}3$ & $321.00$ & $1.38\mathrm{e}5$ & $318.00$ & $318.00$ & $318.00$ \\
neos-950242 & $4.00$ & $4.00$ & $4.00$ & $4.00$ & $4.00$ & $4.00$ & $4.00$ \\
\midrule
neos-957323 & $-237.75$ & $-237.75$ & $-237.75$ & $-229.72$ & $-237.76$ & $-237.76$ & $-237.76$ \\
neos-960392 & $-236.00$ & $-36.00$ & $-238.00$ & $0.00$ & $-238.00$ & $-238.00$ & $-238.00$ \\
neos17 & $0.35$ & $0.35$ & $0.36$ & $0.41$ & $0.15$ & $0.15$ & $0.15$ \\
neos5 & $15.00$ & $15.00$ & $15.00$ & $15.08$ & $15.00$ & $15.00$ & $15.00$ \\
neos8 & $-3.72\mathrm{e}3$ & $-3.72\mathrm{e}3$ & $-3.52\mathrm{e}3$ & $-3.52\mathrm{e}3$ & $-3.72\mathrm{e}3$ & $-3.72\mathrm{e}3$ & $-3.72\mathrm{e}3$ \\
net12 & $214.00$ & $214.00$ & $214.00$ & $214.00$ & $214.00$ & $296.00$ & $214.00$ \\
netdiversion & $398.00$ & $440.00$ & $514.00$ & $525.00$ & $242.00$ & $382.00$ & $242.00$ \\
nexp-150-20-8-5 & $253.00$ & $250.00$ & $258.00$ & $262.00$ & $231.00$ & $243.00$ & $231.00$ \\
ns1116954 & $0.00$ & $0.00$ & $0.00$ & $0.00$ & $0.00$ & Inf & $0.00$ \\
ns1208400 & $2.00$ & $2.00$ & $2.00$ & $2.00$ & $2.00$ & $2.00$ & $2.00$ \\
\bottomrule
\end{tabular}
}
{}
\end{table}

\begin{table}
\TABLE
{{Computational results on the MIPLIB test set in terms of feasible objective value (continued).}\label{app:tab:computational.results.miplib.4}}
{\setlength{\tabcolsep}{3pt}
\footnotesize
\begin{tabular}{@{}crrrrrrr@{}}
\toprule
\multicolumn{1}{c}{\multirow{2}{*}{Instance}} & \multicolumn{7}{c}{Final Feasible Objective} \\
\cmidrule(lr){2-8}
 & \multicolumn{1}{c}{PaNGEA} & \multicolumn{1}{c}{\shortstack{PaNGEA w/o\\CPU thread}} & \multicolumn{1}{c}{\shortstack{PaNGEA\\W=1,R=1}} & \multicolumn{1}{c}{Local-MIP} & \multicolumn{1}{c}{\shortstack{Gurobi\\heuristics}} & \multicolumn{1}{c}{\shortstack{Gurobi w/o\\branching}} & \multicolumn{1}{c}{\shortstack{Provided\\baseline}} \\
\midrule
ns1644855 & $-791.33$ & $-933.00$ & $-302.79$ & $-46.71$ & $-1.52\mathrm{e}3$ & $-1.52\mathrm{e}3$ & $-1.52\mathrm{e}3$ \\
ns1760995 & $-539.11$ & $-536.14$ & $-549.21$ & $-547.28$ & $-549.21$ & $-360.49$ & $-549.21$ \\
ns1830653 & $2.06\mathrm{e}4$ & $2.26\mathrm{e}4$ & $2.06\mathrm{e}4$ & $2.56\mathrm{e}4$ & $2.06\mathrm{e}4$ & $3.46\mathrm{e}4$ & $2.06\mathrm{e}4$ \\
ns1952667 & Inf & Inf & Inf & Inf & $0.00$ & Inf & $0.00$ \\
nu25-pr12 & $5.39\mathrm{e}4$ & $5.39\mathrm{e}4$ & $5.39\mathrm{e}4$ & $5.39\mathrm{e}4$ & $5.39\mathrm{e}4$ & $5.39\mathrm{e}4$ & $5.39\mathrm{e}4$ \\
nursesched-medium-hint03 & $2.41\mathrm{e}3$ & $2.19\mathrm{e}3$ & $4.01\mathrm{e}3$ & $8.35\mathrm{e}3$ & $118.00$ & $553.00$ & $115.00$ \\
nursesched-sprint02 & $74.00$ & $88.00$ & $65.00$ & $79.00$ & $58.00$ & $58.00$ & $58.00$ \\
nw04 & $1.73\mathrm{e}4$ & $1.70\mathrm{e}4$ & $1.81\mathrm{e}4$ & $2.11\mathrm{e}4$ & $1.69\mathrm{e}4$ & $1.69\mathrm{e}4$ & $1.69\mathrm{e}4$ \\
opm2-z10-s4 & $-2.81\mathrm{e}4$ & $-2.87\mathrm{e}4$ & $-2.91\mathrm{e}4$ & $-2.75\mathrm{e}4$ & $-3.27\mathrm{e}4$ & $-3.24\mathrm{e}4$ & $-3.33\mathrm{e}4$ \\
p200x1188c & $1.51\mathrm{e}4$ & $1.51\mathrm{e}4$ & $1.51\mathrm{e}4$ & $1.51\mathrm{e}4$ & $1.51\mathrm{e}4$ & $1.51\mathrm{e}4$ & $1.51\mathrm{e}4$ \\
\midrule
peg-solitaire-a3 & Inf & Inf & Inf & Inf & $1.00$ & Inf & $1.00$ \\
pg & $-7.96\mathrm{e}3$ & $-7.87\mathrm{e}3$ & $-6.40\mathrm{e}3$ & $-5.52\mathrm{e}3$ & $-8.67\mathrm{e}3$ & $-8.65\mathrm{e}3$ & $-8.67\mathrm{e}3$ \\
pg5\_34 & $-1.42\mathrm{e}4$ & $-1.42\mathrm{e}4$ & $-1.37\mathrm{e}4$ & $-1.16\mathrm{e}4$ & $-1.43\mathrm{e}4$ & $-1.43\mathrm{e}4$ & $-1.43\mathrm{e}4$ \\
physiciansched3-3 & $2.97\mathrm{e}6$ & $3.09\mathrm{e}6$ & $3.02\mathrm{e}6$ & $3.12\mathrm{e}6$ & $2.65\mathrm{e}6$ & $3.42\mathrm{e}6$ & $2.62\mathrm{e}6$ \\
physiciansched6-2 & $4.93\mathrm{e}4$ & $4.93\mathrm{e}4$ & $4.93\mathrm{e}4$ & $4.93\mathrm{e}4$ & $4.93\mathrm{e}4$ & $4.93\mathrm{e}4$ & $4.93\mathrm{e}4$ \\
piperout-08 & $1.25\mathrm{e}5$ & $1.42\mathrm{e}5$ & $1.25\mathrm{e}5$ & $1.25\mathrm{e}5$ & $1.25\mathrm{e}5$ & $1.25\mathrm{e}5$ & $1.25\mathrm{e}5$ \\
piperout-27 & $8.12\mathrm{e}3$ & $8.12\mathrm{e}3$ & $8.12\mathrm{e}3$ & $8.12\mathrm{e}3$ & $8.12\mathrm{e}3$ & $8.12\mathrm{e}3$ & $8.12\mathrm{e}3$ \\
pk1 & $23.00$ & $18.00$ & $22.00$ & $28.00$ & $11.00$ & $22.00$ & $11.00$ \\
proteindesign121hz512p9 & $1.50\mathrm{e}3$ & $1.51\mathrm{e}3$ & $1.49\mathrm{e}3$ & $1.51\mathrm{e}3$ & $1.48\mathrm{e}3$ & $1.49\mathrm{e}3$ & $1.47\mathrm{e}3$ \\
proteindesign122trx11p8 & $1.76\mathrm{e}3$ & $1.76\mathrm{e}3$ & $1.75\mathrm{e}3$ & $1.76\mathrm{e}3$ & $1.75\mathrm{e}3$ & $1.76\mathrm{e}3$ & $1.75\mathrm{e}3$ \\
\midrule
qap10 & $340.00$ & $418.00$ & $368.00$ & $376.00$ & $340.00$ & $340.00$ & $340.00$ \\
radiationm18-12-05 & $1.76\mathrm{e}4$ & $1.76\mathrm{e}4$ & $1.76\mathrm{e}4$ & $1.76\mathrm{e}4$ & $1.76\mathrm{e}4$ & $1.76\mathrm{e}4$ & $1.76\mathrm{e}4$ \\
radiationm40-10-02 & $1.55\mathrm{e}5$ & $1.55\mathrm{e}5$ & $1.55\mathrm{e}5$ & $1.55\mathrm{e}5$ & $1.55\mathrm{e}5$ & $1.55\mathrm{e}5$ & $1.55\mathrm{e}5$ \\
rail01 & Inf & Inf & Inf & Inf & Inf & Inf & $-70.57$ \\
rail02 & Inf & Inf & Inf & Inf & Inf & Inf & $-200.45$ \\
rail507 & $176.00$ & $175.00$ & $179.00$ & $190.00$ & $174.00$ & $174.00$ & $174.00$ \\
ran14x18-disj-8 & $4.06\mathrm{e}3$ & $4.02\mathrm{e}3$ & $4.19\mathrm{e}3$ & $4.42\mathrm{e}3$ & $3.71\mathrm{e}3$ & $3.77\mathrm{e}3$ & $3.71\mathrm{e}3$ \\
rd-rplusc-21 & Inf & Inf & Inf & Inf & $1.65\mathrm{e}5$ & Inf & $1.65\mathrm{e}5$ \\
reblock115 & $-3.57\mathrm{e}7$ & $-3.56\mathrm{e}7$ & $-3.56\mathrm{e}7$ & $-3.54\mathrm{e}7$ & $-3.68\mathrm{e}7$ & $-3.68\mathrm{e}7$ & $-3.68\mathrm{e}7$ \\
rmatr100-p10 & $433.00$ & $480.00$ & $459.00$ & $425.00$ & $423.00$ & $429.00$ & $423.00$ \\
\midrule
rmatr200-p5 & $5.46\mathrm{e}3$ & $5.67\mathrm{e}3$ & $5.39\mathrm{e}3$ & $5.31\mathrm{e}3$ & $4.52\mathrm{e}3$ & $4.64\mathrm{e}3$ & $4.52\mathrm{e}3$ \\
rocI-4-11 & $-5.05\mathrm{e}6$ & $-4.04\mathrm{e}6$ & $-5.05\mathrm{e}6$ & $-5.05\mathrm{e}6$ & $-5.05\mathrm{e}6$ & $-4.04\mathrm{e}6$ & $-6.02\mathrm{e}6$ \\
rocII-5-11 & $-5.66$ & $-1.66$ & $-5.66$ & $-3.50$ & $-5.68$ & Inf & $-6.68$ \\
rococoB10-011000 & $2.56\mathrm{e}4$ & $2.98\mathrm{e}4$ & $2.40\mathrm{e}4$ & $2.64\mathrm{e}4$ & $1.94\mathrm{e}4$ & $2.14\mathrm{e}4$ & $1.94\mathrm{e}4$ \\
rococoC10-001000 & $1.40\mathrm{e}4$ & $1.30\mathrm{e}4$ & $1.41\mathrm{e}4$ & $1.58\mathrm{e}4$ & $1.15\mathrm{e}4$ & $1.16\mathrm{e}4$ & $1.15\mathrm{e}4$ \\
roi2alpha3n4 & $-50.40$ & $-32.30$ & $-9.68$ & $-57.53$ & $-63.21$ & $-60.85$ & $-63.21$ \\
roi5alpha10n8 & $-9.45$ & $-12.66$ & $-1.74$ & $-0.24$ & $-51.54$ & $-48.68$ & $-52.32$ \\
roll3000 & $1.42\mathrm{e}4$ & $1.45\mathrm{e}4$ & $1.49\mathrm{e}4$ & $1.51\mathrm{e}4$ & $1.29\mathrm{e}4$ & $1.32\mathrm{e}4$ & $1.29\mathrm{e}4$ \\
s100 & $-0.12$ & $-0.12$ & $-0.14$ & $-0.13$ & $-0.17$ & $-0.17$ & $-0.17$ \\
s250r10 & $-0.12$ & $-0.09$ & $-0.12$ & $-0.11$ & $-0.17$ & $-0.17$ & $-0.17$ \\
\midrule
satellites2-40 & $22.00$ & $27.00$ & $-17.00$ & $22.00$ & $-19.00$ & $16.00$ & $-19.00$ \\
satellites2-60-fs & $-16.00$ & $5.00$ & $9.00$ & $-19.00$ & $-19.00$ & $-19.00$ & $-19.00$ \\
savsched1 & $1.60\mathrm{e}4$ & $1.05\mathrm{e}5$ & $1.06\mathrm{e}4$ & $1.64\mathrm{e}4$ & $3.22\mathrm{e}3$ & $3.22\mathrm{e}3$ & $3.22\mathrm{e}3$ \\
sct2 & $-228.89$ & $-228.05$ & $-224.11$ & $-222.58$ & $-230.99$ & $-230.99$ & $-230.99$ \\
seymour & $423.00$ & $424.00$ & $423.00$ & $424.00$ & $423.00$ & $426.00$ & $423.00$ \\
seymour1 & $412.87$ & $413.61$ & $414.67$ & $415.10$ & $410.76$ & $410.84$ & $410.76$ \\
sing326 & $1.06\mathrm{e}7$ & $1.32\mathrm{e}7$ & $2.51\mathrm{e}7$ & $1.52\mathrm{e}8$ & $7.75\mathrm{e}6$ & $7.75\mathrm{e}6$ & $7.75\mathrm{e}6$ \\
sing44 & $8.77\mathrm{e}6$ & $9.69\mathrm{e}6$ & $1.19\mathrm{e}7$ & $1.35\mathrm{e}8$ & $8.13\mathrm{e}6$ & $8.13\mathrm{e}6$ & $8.13\mathrm{e}6$ \\
snp-02-004-104 & $1.19\mathrm{e}9$ & $1.31\mathrm{e}9$ & $6.57\mathrm{e}8$ & $8.11\mathrm{e}9$ & $5.87\mathrm{e}8$ & $5.87\mathrm{e}8$ & $5.87\mathrm{e}8$ \\
sorrell3 & $-16.00$ & $-16.00$ & $-16.00$ & $-15.00$ & $-16.00$ & $-15.00$ & $-16.00$ \\
\bottomrule
\end{tabular}
}
{}
\end{table}

\begin{table}
\TABLE
{{Computational results on the MIPLIB test set in terms of feasible objective value (continued).}\label{app:tab:computational.results.miplib.5}}
{\setlength{\tabcolsep}{3pt}
\footnotesize
\begin{tabular}{@{}crrrrrrr@{}}
\toprule
\multicolumn{1}{c}{\multirow{2}{*}{Instance}} & \multicolumn{7}{c}{Final Feasible Objective} \\
\cmidrule(lr){2-8}
 & \multicolumn{1}{c}{PaNGEA} & \multicolumn{1}{c}{\shortstack{PaNGEA w/o\\CPU thread}} & \multicolumn{1}{c}{\shortstack{PaNGEA\\W=1,R=1}} & \multicolumn{1}{c}{Local-MIP} & \multicolumn{1}{c}{\shortstack{Gurobi\\heuristics}} & \multicolumn{1}{c}{\shortstack{Gurobi w/o\\branching}} & \multicolumn{1}{c}{\shortstack{Provided\\baseline}} \\
\midrule
sp150x300d & $69.00$ & $69.00$ & $69.00$ & $69.00$ & $69.00$ & $69.00$ & $69.00$ \\
sp97ar & $6.90\mathrm{e}8$ & $7.06\mathrm{e}8$ & $7.32\mathrm{e}8$ & $1.08\mathrm{e}9$ & $6.61\mathrm{e}8$ & $6.63\mathrm{e}8$ & $6.61\mathrm{e}8$ \\
sp98ar & $5.59\mathrm{e}8$ & $5.45\mathrm{e}8$ & $5.68\mathrm{e}8$ & $8.29\mathrm{e}8$ & $5.30\mathrm{e}8$ & $5.30\mathrm{e}8$ & $5.30\mathrm{e}8$ \\
splice1k1 & $-394.00$ & $-394.00$ & $-394.00$ & $-394.00$ & $-394.00$ & $-338.00$ & $-394.00$ \\
square41 & $64.00$ & $56.00$ & $123.00$ & $121.00$ & $15.00$ & $15.00$ & $15.00$ \\
square47 & $76.00$ & $210.00$ & Inf & Inf & $20.00$ & $20.00$ & $16.00$ \\
supportcase10 & $10.00$ & $16.00$ & $8.00$ & $10.00$ & $9.00$ & $9.00$ & $7.00$ \\
supportcase12 & $-7.19\mathrm{e}3$ & $-2.86\mathrm{e}3$ & $-2.96\mathrm{e}3$ & $-154.14$ & $-7.56\mathrm{e}3$ & $-7.55\mathrm{e}3$ & $-7.56\mathrm{e}3$ \\
supportcase18 & $50.00$ & $50.00$ & $50.00$ & $50.00$ & $48.00$ & $49.00$ & $48.00$ \\
supportcase19 & Inf & Inf & Inf & Inf & Inf & Inf & $1.27\mathrm{e}7$ \\
\midrule
supportcase22 & Inf & Inf & Inf & Inf & Inf & Inf & $110.00$ \\
supportcase26 & $1.75\mathrm{e}3$ & $1.75\mathrm{e}3$ & $1.75\mathrm{e}3$ & $1.75\mathrm{e}3$ & $1.75\mathrm{e}3$ & $1.93\mathrm{e}3$ & $1.75\mathrm{e}3$ \\
supportcase33 & $-270.00$ & $-200.00$ & $-270.00$ & $-250.00$ & $-345.00$ & $-315.00$ & $-345.00$ \\
supportcase40 & $2.55\mathrm{e}4$ & $2.56\mathrm{e}4$ & $2.46\mathrm{e}4$ & $3.51\mathrm{e}4$ & $2.43\mathrm{e}4$ & $2.48\mathrm{e}4$ & $2.43\mathrm{e}4$ \\
supportcase42 & $8.00$ & $8.00$ & $8.00$ & $40.00$ & $7.76$ & $8.00$ & $7.76$ \\
supportcase6 & $6.24\mathrm{e}4$ & $6.24\mathrm{e}4$ & $6.21\mathrm{e}4$ & $7.25\mathrm{e}4$ & $5.19\mathrm{e}4$ & $5.19\mathrm{e}4$ & $5.19\mathrm{e}4$ \\
supportcase7 & $-1.01\mathrm{e}3$ & $-983.97$ & $-595.18$ & $-628.74$ & $-1.13\mathrm{e}3$ & $-1.13\mathrm{e}3$ & $-1.13\mathrm{e}3$ \\
swath1 & $386.71$ & $386.71$ & $457.32$ & $876.91$ & $379.07$ & $382.89$ & $379.07$ \\
swath3 & $472.41$ & $450.27$ & $495.24$ & $898.52$ & $397.76$ & $401.69$ & $397.76$ \\
tbfp-network & $81.17$ & $80.29$ & $66.48$ & $76.31$ & $24.16$ & $25.29$ & $24.16$ \\
\midrule
thor50dday & $1.32\mathrm{e}5$ & $1.44\mathrm{e}5$ & $8.51\mathrm{e}4$ & $2.05\mathrm{e}5$ & $4.04\mathrm{e}4$ & $4.04\mathrm{e}4$ & $4.04\mathrm{e}4$ \\
timtab1 & $8.60\mathrm{e}5$ & $8.97\mathrm{e}5$ & $9.24\mathrm{e}5$ & $1.00\mathrm{e}6$ & $7.65\mathrm{e}5$ & $8.78\mathrm{e}5$ & $7.65\mathrm{e}5$ \\
tr12-30 & $1.50\mathrm{e}5$ & $1.52\mathrm{e}5$ & $1.56\mathrm{e}5$ & $1.62\mathrm{e}5$ & $1.31\mathrm{e}5$ & $1.31\mathrm{e}5$ & $1.31\mathrm{e}5$ \\
traininstance2 & $7.48\mathrm{e}4$ & $7.66\mathrm{e}4$ & $7.43\mathrm{e}4$ & $7.50\mathrm{e}4$ & $7.18\mathrm{e}4$ & $7.86\mathrm{e}4$ & $7.18\mathrm{e}4$ \\
traininstance6 & $2.84\mathrm{e}4$ & $2.87\mathrm{e}4$ & $2.84\mathrm{e}4$ & $2.83\mathrm{e}4$ & $2.83\mathrm{e}4$ & $3.29\mathrm{e}4$ & $2.83\mathrm{e}4$ \\
trento1 & $2.52\mathrm{e}7$ & $5.55\mathrm{e}7$ & $3.53\mathrm{e}7$ & $1.26\mathrm{e}8$ & $5.19\mathrm{e}6$ & $5.65\mathrm{e}6$ & $5.19\mathrm{e}6$ \\
triptim1 & Inf & Inf & Inf & Inf & $22.87$ & $22.87$ & $22.87$ \\
uccase12 & $1.35\mathrm{e}4$ & $1.15\mathrm{e}4$ & $1.55\mathrm{e}4$ & $9.50\mathrm{e}5$ & $1.15\mathrm{e}4$ & $1.15\mathrm{e}4$ & $1.15\mathrm{e}4$ \\
uccase9 & $1.25\mathrm{e}4$ & $2.76\mathrm{e}4$ & $2.47\mathrm{e}4$ & $2.57\mathrm{e}6$ & $1.10\mathrm{e}4$ & $1.11\mathrm{e}4$ & $1.10\mathrm{e}4$ \\
uct-subprob & $334.00$ & $334.00$ & $347.00$ & $353.00$ & $314.00$ & $341.00$ & $314.00$ \\
\midrule
unitcal\_7 & $2.03\mathrm{e}7$ & $2.04\mathrm{e}7$ & $2.16\mathrm{e}7$ & $2.43\mathrm{e}7$ & $1.96\mathrm{e}7$ & $1.96\mathrm{e}7$ & $1.96\mathrm{e}7$ \\
var-smallemery-m6j6 & $-142.03$ & $-146.19$ & $-136.31$ & $0.00$ & $-149.38$ & $-147.94$ & $-149.38$ \\
wachplan & $-8.00$ & $-8.00$ & $-8.00$ & $-8.00$ & $-8.00$ & $-8.00$ & $-8.00$ \\
\bottomrule
\end{tabular}
}
{}
\end{table}

\FloatBarrier

\section{Integration of PaNGEA and Branch-and-Bound} \label{app:bnb}
Our prototype branch-and-bound code maintains a persistent branch-and-bound tree alongside PaNGEA, starting from the root node, as described in Algorithm~\ref{alg:pangea_bnb}. 

The branch-and-bound part follows a standard structure: 
(1) \emph{node selection}: because multiple tree nodes can be processed in the same batch, we can smoothly balance the search for a good solution and improving the lower bound. Hence, for a given number of nodes to process in parallel $B_{\texttt{B\&B}}$, we allocate half of the budget to nodes with the smallest lower bounds and the other half to nodes with the largest lower bounds;
(2) \emph{node processing}: at each iteration, we solve the linear relaxations of the nodes with cuPDLP, from which we compute dual lower bounds that remain valid even if the algorithm terminates before reaching the desired accuracy; 
(3) \emph{branching}: we branch on the most-fractional binary variable (integer variable when all binary variables are fixed), followed by bound propagation; and (4) \emph{pruning}: nodes are removed when the node relaxation is found infeasible, when the computed lower bound exceeds $\texttt{bestUB}$, or when bound propagation detects a conflict. 

The PaNGEA part follows Section~\ref{sec:alg}, supplying $\texttt{bestUB}$ to the branch-and-bound part. 

In our implementation, we use $B_{\texttt{B\&B}} = B_{\texttt{PaNGEA}} = B/2$.

\begin{algorithm}[t]
\small
\caption{{PaNGEA integrated into branch-and-bound}}
\label{alg:pangea_bnb}
\KwIn{Problem $\mathcal{P}$, batch size $B = B_{\texttt{B\&B}} + B_{\texttt{PaNGEA}}$}
Initialize PaNGEA nodes from the root and $\texttt{bestUB}$ as in Section~\ref{sec:alg}\;
Initialize the set of active branch-and-bound nodes $\mathcal{F}$ with the root node\;
\While{$\mathcal{F}\neq\emptyset$ and time limit is not reached}{
    \textbf{Prune} branch-and-bound nodes whose lower bound $\geq$ $\texttt{bestUB}$\;
    \textbf{Select} up to $B_{\texttt{B\&B}}$ nodes  $\mathcal{N}\subseteq\mathcal{F}$: half with smallest lower bounds, half with largest\;
    Update $\mathcal{F}\leftarrow\mathcal{F}\setminus\mathcal{N}$\;
    Generate $B_{\texttt{PaNGEA}}$ new PaNGEA nodes, $\mathcal{H}$ \;
    \textbf{Solve} the linear relaxations of $\mathcal{N}\cup\mathcal{H}$ in a batch\;
    Update the lower bounds for nodes in $\mathcal{N}$\;
    Run batched local search for $\mathcal{H}$; Update $\texttt{bestUB}$\;
    \textbf{Branch} on the nodes in $\mathcal{N}$; 
    Propagate bounds;
    Drop infeasible nodes and add the remaining into $\mathcal{F}$ \;
    Select PaNGEA nodes as in Section~\ref{sec:alg.node.selection}.
}
\end{algorithm}

\end{APPENDICES}

\FloatBarrier
\begingroup \parindent 0pt \parskip 0.0ex \def\enotesize{\normalsize} \theendnotes \endgroup

\end{document}